\RequirePackage{lineno}
\documentclass[aps,prd,twocolumn,showpacs,superscriptaddress,groupedaddress,nofootinbib]{revtex4}

 \usepackage{graphicx}
\usepackage{amssymb}

\usepackage{verbatim}
\usepackage{subfigure}

\usepackage{alphalph}

\newcommand{\dzero}  {D0}
\newcommand{\dca}    {\ensuremath{{\rm dca}}}
\newcommand{\lumi}   {\ensuremath{{L}}}
\newcommand{\pt}     {\ensuremath{p_T}}
\newcommand{\et}     {\ensuremath{E_T}}
\newcommand{\etmiss}{\mbox{\ensuremath{E\kern-0.6em\slash_T}}}

\newcommand{\dr}     {\ensuremath{\Delta R}}
\newcommand{\gev}    {\ensuremath{\rm GeV}}
\newcommand{\ical}   {\ensuremath{\mathcal{I}^{\rm cal}}}
\newcommand{\itrk}   {\ensuremath{\mathcal{I}^{\rm trk}}}

\newcommand{\fbi}    {\ensuremath{\mathrm{ fb^{-1}}}}
\newcommand{\mumu}   {\ensuremath{\mu^+\mu^-}}
\newcommand{\zgmumu} {\ensuremath{Z/\gamma^\star\to \mumu}}
\newcommand{\zmumu}  {\ensuremath{Z\to \mumu}}
\newcommand{\jpsimumu}{\ensuremath{J/\psi\to\mumu}}
\newcommand{\jpsi}{\ensuremath{J/\psi}}
\newcommand{\GeV}{\ensuremath{{\rm GeV}}}

\newcommand{\etadet}{\eta_{\rm {detector}}}
\newcommand{\etacft}{\eta_{\rm {CFT}}}
\newcommand{\CFT}{\rm{CFT}}
\newcommand{\LARM}{L_{\rm{arm}}}

\def\ie{{\it i.e.}}

\begin{document}

\widetext
\hspace{5.2in}\mbox{FERMILAB-PUB-13-272-E}

\title{Muon reconstruction and identification with the Run II D0 detector}

\makeatletter{}\affiliation{LAFEX, Centro Brasileiro de Pesquisas F\'{i}sicas, Rio de Janeiro, Brazil}
\affiliation{Universidade do Estado do Rio de Janeiro, Rio de Janeiro, Brazil}
\affiliation{Universidade Federal do ABC, Santo Andr\'e, Brazil}
\affiliation{University of Science and Technology of China, Hefei, People's Republic of China}
\affiliation{Universidad de los Andes, Bogot\'a, Colombia}
\affiliation{Charles University, Faculty of Mathematics and Physics, Center for Particle Physics, Prague, Czech Republic}
\affiliation{Czech Technical University in Prague, Prague, Czech Republic}
\affiliation{Institute of Physics, Academy of Sciences of the Czech Republic, Prague, Czech Republic}
\affiliation{Universidad San Francisco de Quito, Quito, Ecuador}
\affiliation{LPC, Universit\'e Blaise Pascal, CNRS/IN2P3, Clermont, France}
\affiliation{LPSC, Universit\'e Joseph Fourier Grenoble 1, CNRS/IN2P3, Institut National Polytechnique de Grenoble, Grenoble, France}
\affiliation{CPPM, Aix-Marseille Universit\'e, CNRS/IN2P3, Marseille, France}
\affiliation{LAL, Universit\'e Paris-Sud, CNRS/IN2P3, Orsay, France}
\affiliation{LPNHE, Universit\'es Paris VI and VII, CNRS/IN2P3, Paris, France}
\affiliation{CEA, Irfu, SPP, Saclay, France}
\affiliation{IPHC, Universit\'e de Strasbourg, CNRS/IN2P3, Strasbourg, France}
\affiliation{IPNL, Universit\'e Lyon 1, CNRS/IN2P3, Villeurbanne, France and Universit\'e de Lyon, Lyon, France}
\affiliation{III. Physikalisches Institut A, RWTH Aachen University, Aachen, Germany}
\affiliation{Physikalisches Institut, Universit\"at Freiburg, Freiburg, Germany}
\affiliation{II. Physikalisches Institut, Georg-August-Universit\"at G\"ottingen, G\"ottingen, Germany}
\affiliation{Institut f\"ur Physik, Universit\"at Mainz, Mainz, Germany}
\affiliation{Ludwig-Maximilians-Universit\"at M\"unchen, M\"unchen, Germany}
\affiliation{Panjab University, Chandigarh, India}
\affiliation{Delhi University, Delhi, India}
\affiliation{Tata Institute of Fundamental Research, Mumbai, India}
\affiliation{University College Dublin, Dublin, Ireland}
\affiliation{Korea Detector Laboratory, Korea University, Seoul, Korea}
\affiliation{CINVESTAV, Mexico City, Mexico}
\affiliation{Nikhef, Science Park, Amsterdam, the Netherlands}
\affiliation{Radboud University Nijmegen, Nijmegen, the Netherlands}
\affiliation{Joint Institute for Nuclear Research, Dubna, Russia}
\affiliation{Institute for Theoretical and Experimental Physics, Moscow, Russia}
\affiliation{Moscow State University, Moscow, Russia}
\affiliation{Institute for High Energy Physics, Protvino, Russia}
\affiliation{Petersburg Nuclear Physics Institute, St. Petersburg, Russia}
\affiliation{Instituci\'{o} Catalana de Recerca i Estudis Avan\c{c}ats (ICREA) and Institut de F\'{i}sica d'Altes Energies (IFAE), Barcelona, Spain}
\affiliation{Stockholm University, Stockholm and Uppsala University, Uppsala, Sweden}
\affiliation{Lancaster University, Lancaster LA1 4YB, United Kingdom}
\affiliation{Imperial College London, London SW7 2AZ, United Kingdom}
\affiliation{The University of Manchester, Manchester M13 9PL, United Kingdom}
\affiliation{University of Arizona, Tucson, Arizona 85721, USA}
\affiliation{University of California Riverside, Riverside, California 92521, USA}
\affiliation{Florida State University, Tallahassee, Florida 32306, USA}
\affiliation{Fermi National Accelerator Laboratory, Batavia, Illinois 60510, USA}
\affiliation{University of Illinois at Chicago, Chicago, Illinois 60607, USA}
\affiliation{Northern Illinois University, DeKalb, Illinois 60115, USA}
\affiliation{Northwestern University, Evanston, Illinois 60208, USA}
\affiliation{Indiana University, Bloomington, Indiana 47405, USA}
\affiliation{Purdue University Calumet, Hammond, Indiana 46323, USA}
\affiliation{University of Notre Dame, Notre Dame, Indiana 46556, USA}
\affiliation{Iowa State University, Ames, Iowa 50011, USA}
\affiliation{University of Kansas, Lawrence, Kansas 66045, USA}
\affiliation{Louisiana Tech University, Ruston, Louisiana 71272, USA}
\affiliation{Northeastern University, Boston, Massachusetts 02115, USA}
\affiliation{University of Michigan, Ann Arbor, Michigan 48109, USA}
\affiliation{Michigan State University, East Lansing, Michigan 48824, USA}
\affiliation{University of Mississippi, University, Mississippi 38677, USA}
\affiliation{University of Nebraska, Lincoln, Nebraska 68588, USA}
\affiliation{Rutgers University, Piscataway, New Jersey 08855, USA}
\affiliation{Princeton University, Princeton, New Jersey 08544, USA}
\affiliation{State University of New York, Buffalo, New York 14260, USA}
\affiliation{University of Rochester, Rochester, New York 14627, USA}
\affiliation{State University of New York, Stony Brook, New York 11794, USA}
\affiliation{Brookhaven National Laboratory, Upton, New York 11973, USA}
\affiliation{Langston University, Langston, Oklahoma 73050, USA}
\affiliation{University of Oklahoma, Norman, Oklahoma 73019, USA}
\affiliation{Oklahoma State University, Stillwater, Oklahoma 74078, USA}
\affiliation{Brown University, Providence, Rhode Island 02912, USA}
\affiliation{University of Texas, Arlington, Texas 76019, USA}
\affiliation{Southern Methodist University, Dallas, Texas 75275, USA}
\affiliation{Rice University, Houston, Texas 77005, USA}
\affiliation{University of Virginia, Charlottesville, Virginia 22904, USA}
\affiliation{University of Washington, Seattle, Washington 98195, USA}
\author{V.M.~Abazov} \affiliation{Joint Institute for Nuclear Research, Dubna, Russia}
\author{B.~Abbott} \affiliation{University of Oklahoma, Norman, Oklahoma 73019, USA}
\author{B.S.~Acharya} \affiliation{Tata Institute of Fundamental Research, Mumbai, India}
\author{M.~Adams} \affiliation{University of Illinois at Chicago, Chicago, Illinois 60607, USA}
\author{T.~Adams} \affiliation{Florida State University, Tallahassee, Florida 32306, USA}
\author{J.P.~Agnew} \affiliation{The University of Manchester, Manchester M13 9PL, United Kingdom}
\author{G.D.~Alexeev} \affiliation{Joint Institute for Nuclear Research, Dubna, Russia}
\author{G.~Alkhazov} \affiliation{Petersburg Nuclear Physics Institute, St. Petersburg, Russia}
\author{A.~Alton$^{a}$} \affiliation{University of Michigan, Ann Arbor, Michigan 48109, USA}
\author{M.~Arthaud} \affiliation{CEA, Irfu, SPP, Saclay, France}
% ...
\collaboration{The D0 Collaboration\footnote{with visitors from
$^{a}$Augustana College, Sioux Falls, SD, USA,
$^{b}$The University of Liverpool, Liverpool, UK,
$^{c}$DESY, Hamburg, Germany,
$^{d}$Universidad Michoacana de San Nicolas de Hidalgo, Morelia, Mexico
$^{e}$SLAC, Menlo Park, CA, USA,
$^{f}$University College London, London, UK,
$^{g}$Centro de Investigacion en Computacion - IPN, Mexico City, Mexico,
$^{h}$Universidade Estadual Paulista, S\~ao Paulo, Brazil,
$^{i}$Karlsruher Institut f\"ur Technologie (KIT) - Steinbuch Centre for Computing (SCC)
and
$^{j}$Office of Science, U.S. Department of Energy, Washington, D.C. 20585, USA.
}} \noaffiliation
\vskip 0.25cm
 
\date{July 17, 2013 }

\begin{abstract}
\makeatletter{}

We present an overview of the muon reconstruction and identification methods
employed by the  \dzero\ collaboration to analyze the Run~II~(2001--2011)
$p\bar p$ data of the Fermilab Tevatron collider at $\sqrt s= 1.96$~TeV.
We discuss the performance of these methods, how it is measured using \dzero\ data, and how 
it is properly  modeled by the \dzero\ simulation program.
In its pseudorapidity acceptance, $|\eta|<2$, the muon system identifies
high-\pt\ muons ($\pt\gtrsim 10$ GeV) with efficiencies ranging from 72\%  to 89\%.
Muons tracks are reconstructed in the \dzero\ central tracking system
with efficiencies ranging from 85\% to 92\% and with
a  typical relative momentum resolution of  10\% for  $\pt=40~\gev$.
Isolation criteria  reject multijet background  with efficiencies of 87\% to 99\%.
 
\end{abstract}


\maketitle

\renewcommand{\thefootnote}{\alph{footnote}}

\section{Introduction}
\label{sec:introduction}
\makeatletter{}Muon reconstruction and identification are cornerstones of the \dzero\ Run~II physics program at the Fermilab Tevatron $p\bar p$ collider.
Muons of high transverse momentum ($\pt \gtrsim 10$~GeV)
with no extra calorimeter or tracking activity around them
are a signature of electroweak boson decays ($W\to\mu\nu$, $Z/\gamma^\star\to\mu\mu$),  allowing the study of electroweak physics, top quark physics ($t\to Wb$),
and Higgs boson physics ($WH$, $ZH$ production, $H\to WW$ decay).
Lower-\pt\ muons,
resulting for example from the semi-leptonic decays of heavy flavor quarks or from $J/\psi \to \mu\mu$ decay,
give access to a rich flavor physics program.

In this article, we present an overview of the muon reconstruction and identification methods employed by the
\dzero\ experiment to analyze the $10$~fb$^{-1}$ of high quality data collected
at $\sqrt s= 1.96$~TeV
between 2001 and 2011.
We also present their performance in terms of efficiency, momentum resolution, and background rejection.
The class of analyses dedicated to the heavy-flavor physics program are based on low-\pt\ muons and
do not require a determination of absolute muon identification efficiencies since they are generally normalized to a well-known semileptonic decay process.
This article focuses therefore on muon reconstruction and identification performance for high-$\pt$ muons.
As for \dzero, the physics programs of other hadron collider experiments rely on the capability
to reconstruct and identify muons, as discussed for example in Refs.~\cite
{ATLAS:2011hga,ATLAS:2011yga,Ginsburg:2004fa,Chatrchyan:2012xi}.

This article is structured as follows. After an overview of the \dzero\ Run~II detector and its muon system,
we briefly describe the muon reconstruction algorithms and the muon identification criteria.
We then discuss their performance and how we measure that performance  with the \dzero\ data. We also
 briefly discuss the  background estimate for the muon identification
and present the methods employed to correct the simulation for better agreement with the data.

\section{Overview of the D0 detector and its muon system}
\label{sec:overview}
\makeatletter{}

\begin{figure*}[!]
\begin{center}
\includegraphics[width=1\textwidth]{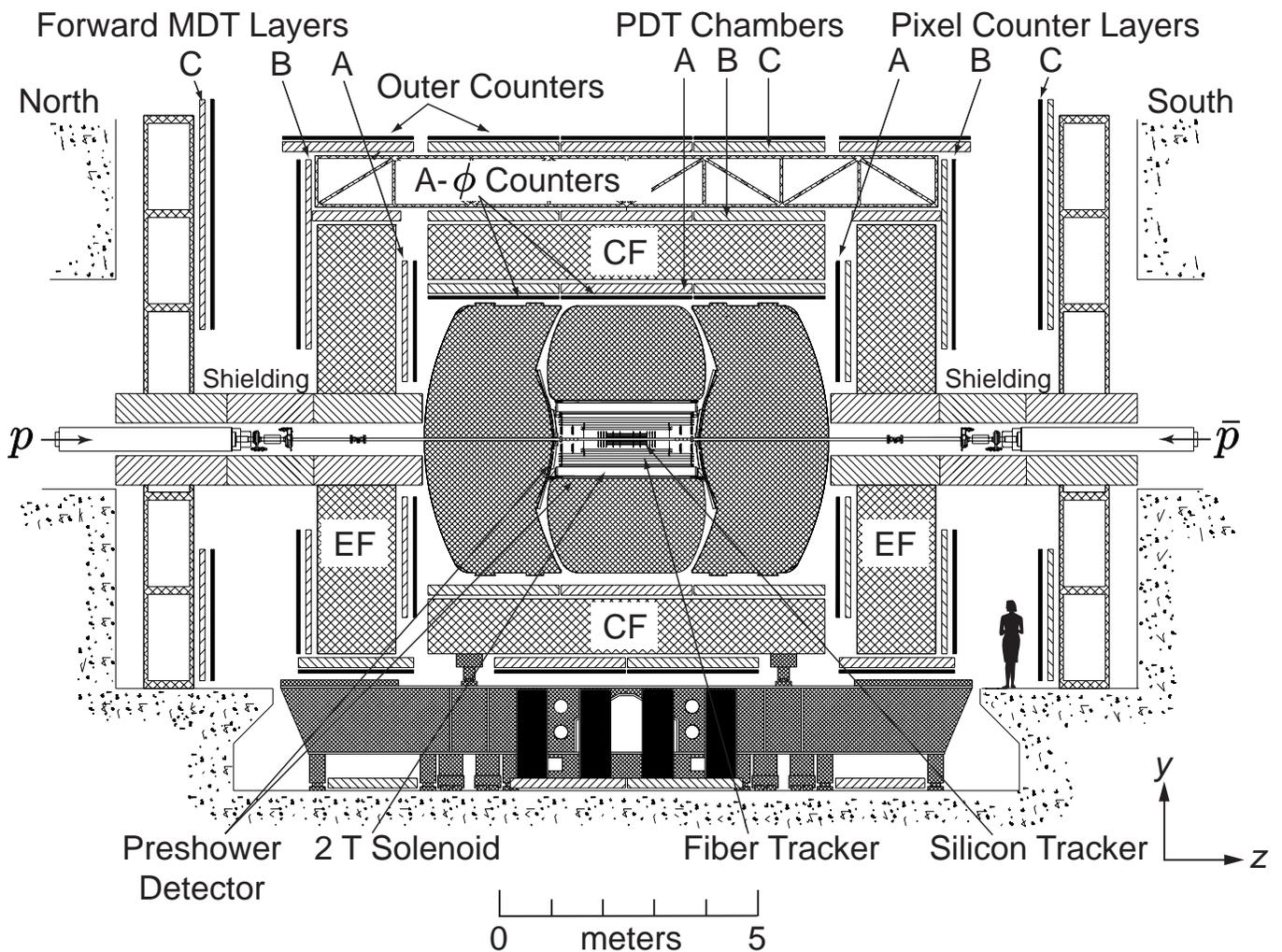}
\end{center}
\caption{
Cross sectional side view of the {\dzero} detector at the beginning of  Run~II.
The abbreviations CF, EF, MDT, and PDT stand for
central toroid,  end toroids,  mini drift tubes, and  proportional drift tubes, respectively.
}\label{d0detector}
\end{figure*}

Figure~\ref{d0detector} presents a cross sectional view of the Run~II  {\dzero}  detector~\cite{upgraded_d0detector,Ahmed20118,Angstadt:2009ie,Abolins:2007yz}.
The detectors surrounding the interaction region include a silicon microstrip detector (SMT) for precision tracking of charged 
particles and determination of the primary vertex and secondary decay vertices, a scintillating fiber tracker (CFT) for precise track reconstruction, 
a 2~T solenoidal magnet for momentum determination of charged particles,
preshower detectors, and the liquid-argon uranium 
calorimeters  with electromagnetic, fine, and coarse hadronic sections.
The muon system, 
which is described in detail in Ref.~\cite{muon_system},
resides outside the calorimeter system
and its main components are identified in Fig.~\ref{d0detector}.
The identification of muons relies mainly on the muon system;
however,
other {\dzero} components, namely the solenoid, CFT, SMT and calorimeter, are used to measure the muon momentum and 
estimate whether a muon is isolated, \ie, there is no extra calorimeter or tracking activity around the particle.

The \dzero\ coordinate system is right-handed, with
the $z$-axis pointing in the direction of the Tevatron proton beam
and the $y$-axis pointing upwards.
The angles $\phi$ and $\theta$ are the azimuthal and polar angles relative to the $x$- and $z$-axes, respectively. We also use as an angular variable the pseudorapidity defined by $\eta=\ln[\tan(\theta/2)]$.

The Run II  muon system  design 
is based on the extensive experience obtained with the Run~I (1992--1996)  muon spectrometer~\cite{d0detector} consisting of 
two muon systems: the wide angle muon system and the small angle muon system. The wide angle system, covering the pseudorapidity 
region $|\eta| \lesssim 2.0$, consisted of proportional drift tubes (PDTs) and three large iron toroidal magnets: a central toroid (CF) 
and two end toroids (EFs). The small angle system, covering $2.0 \lesssim |\eta| \lesssim 3.0$, consisted of a set of drift tube 
planes and two small iron toroids.
For Run~II, the small angle system was completely removed and the small angle magnets were replaced by new shielding assemblies.
In the forward region ($1.0 \lesssim |\eta| \lesssim 2.0$), the PDT chambers were replaced by a new tracking system.
The three main toroids in the  muon magnet system, the CF and the two EFs, were not changed, except that the current 
was reduced by 40\% yielding substantial operating  cost savings.
As the operating point of Run~I was close to the magnetic saturation in iron,
the reduced current resulted in a 6\% lower magnetic field of 1.8~T.
This small change did not degrade the overall performance, since the addition of the solenoid for Run~II provided precision measurements of charged particle momenta close
to the interaction region.

The  Run II muon system consists of one layer of muon detectors before the toroidal magnet and two similar layers of 
detectors after the magnet. This provides the ability to reconstruct and measure the muon track parameters, including the
muon momentum. The three layers of the muon system are referred to as A, B, and C, as 
indicated in Fig.~\ref{d0detector}.
For the purposes of triggering and muon track reconstruction, a system of fast scintillation counters with time resolution of 
$\sigma(t) \approx 2$~ns is used in Run II. 
In the central muon system ($|\eta| \lesssim 1.0$)
there are 630 scintillation counters in the A-layer (referred to as A$\phi$ counters),
with an angular segmentation of   79~mrad in $\phi$, 
and 372 counters in the C-layer (referred to as outer counters). 
In the forward region, a total of 4214 scintillation counters (referred to as pixel counters) are used in the A, B, and C layers, 
providing three independent coordinate and time measurements along muon tracks. They
have a segmentation of approximately 0.1 in $\eta$ and approximately 79~mrad in $\phi$.

The muon system tracking detectors consist of PDTs in the central rapidity region ($|\eta| \lesssim 1.0$) and mini drift tubes (MDTs) 
in the forward region ($1.0 \lesssim |\eta| \lesssim 2.0$).
Both PDTs and MDTs are installed in the three layers, A, B, and C (see Fig.~\ref{d0detector}),
which consist of  4, 3, and 3 detection planes, respectively (except the bottom A-layer PDTs which have three planes).
The PDTs and MDTs
provide high-accuracy coordinate 
measurements with a resolution of approximately 1~mm in the direction perpendicular to the sensitive wires which are arranged parallel to the toroidal field lines.
The PDTs have also Vernier pads~\cite{Brown:1989em}, but only those of layer A are read out to determine the position of the muon hits along the wires.

\begin{figure}[!]
\begin{center}
\includegraphics[width=0.40\textwidth]{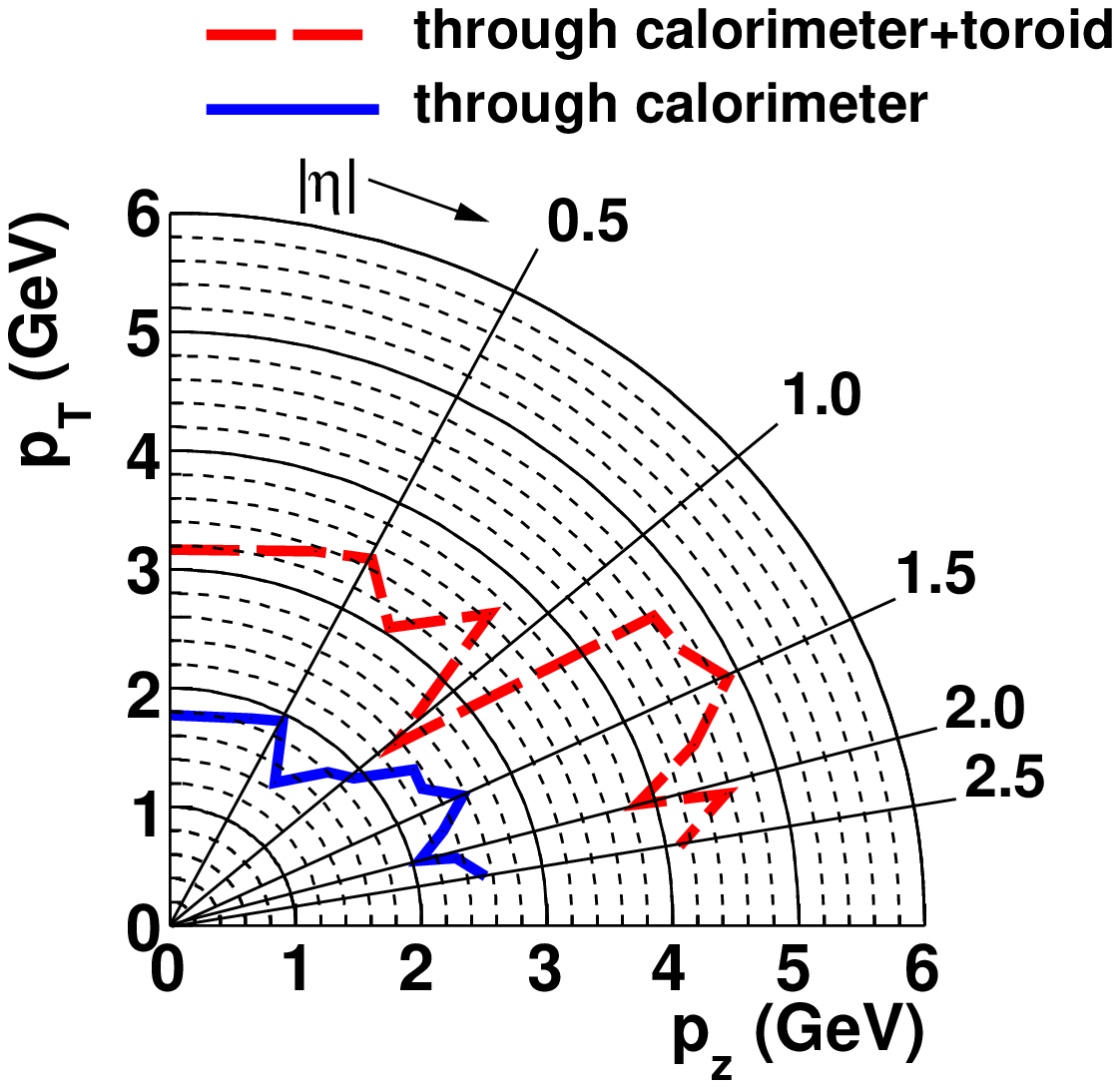}
\end{center}
\caption{ [color online]
Minimum value of muon momenta required
to pass the calorimeter only and both the calorimeter and the toroids,  as functions of pseudorapidity.}
\label{fig:eloss}
\end{figure}

Because of energy loss in the calorimeter, a muon produced in the $p\bar p$ interaction region
must have a minimum energy of 2 to 2.5~GeV,
depending on $|\eta|$, to pass through the calorimeter and reach layer A of the muon system,
as shown in Fig.~\ref{fig:eloss}.
Similarly, the muon energy  must be at least 3 to 5~GeV  to pass through both the uranium calorimeter and
the iron toroid to reach all instrumented layers of the muon system.
 
\section{Muon reconstruction}
\label{sec:reconstruction}
\makeatletter{}
To reconstruct muon trajectories, 
the same algorithm is used in the forward and central regions.
A list of hits from the muon detector is first built.
These hits are associated to form muon track segments,
which are then used to form
tracks in the muon system, called local tracks.
The local tracks and the segments not used in the construction of local tracks
are generically  called local muons.
In a last step the local muons are 
matched to tracks
reconstructed in the central tracking system.
The details of these different steps are discussed below.

\subsection{Hit finding}

For the forward system, the $\phi$ coordinate is determined by the
scintillation counters. As each counter covers 79~mrad, 
the $\phi$ resolution in each layer is  approximately $79/\sqrt {12}= 23$~mrad.

For the central system, the $\phi$ coordinate is determined by both
the central counters and the PDTs.
Similarly to the forward scintillation counters, the measurement of $\phi$ in the central A-layer counters has a resolution of  approximately 23~mrad.
In each of the three layers, the wires of the PDTs are ganged together in pairs
at one end of each chamber and read out by electronics located at the
other end of the chamber.
To obtain the $\phi$ coordinate, the PDTs measure the hit position along the wire
using the time difference of the signal arrival between each end of the pairs.
After correcting for a non-linearity in the response for
those hits with a time difference above 75\% of the maximum value,
position resolutions per wire of 15~cm are obtained. For the
A-layer only, the Vernier pads are
combined with the time division
and scintillator position, which resolve the pad ambiguity and reduce the
position resolution to about 2~mm, which corresponds to about $7$~mrad. 

The time of passage of a particle in a layer of the muon system is given by
the corresponding scintillator time.
Each scintillator time is adjusted based on the path length from
the \dzero\ detector center to the scintillating element,
so that a particle produced in a 
 $p\bar p$ collision  and traveling at the speed of light
will have an expected time of zero.
Given the size of the central counters, a correction is needed
to account for the time it takes for the light produced in the
scintillator to reach the phototube. It is found to be about
0.05~ns$\cdot$cm$^{-1}$ for all counters, consistent with the speed of light in the scintillating material. Including the correction improves the
time resolution from 2.1~ns to 2.0~ns for the A-layer counters, and
from 4.5~ns to 3.8~ns for the large C-layer counters on the sides and top.
The bottom B- and C-layer counter resolutions are improved from
3.7~ns and 3.5~ns to 3.1~ns and 2.5~ns, respectively.

For both the central and forward muon systems,
the drift time of the wire chambers is used to measure
the impact position of a muon perpendicular to the wire. 
The time-to-distance relation is not linear
and depends on the muon incidence angle
because of the geometry of the drift tubes.
In the first iteration,  the muon track is assumed to be orthogonal to the drift chambers, then
the segment direction is used.
Both the left and right ambiguity in the drift time are treated
as individual hits.
The position resolution of both the MDT and PDT systems is approximately 1~mm,
primarily due to the position accuracy of the wire.

\subsection{Segment reconstruction}

In a first step, 2D segments in the
deviation plane, \ie, the plane orthogonal to the toroidal magnetic field and the wires,
are found in each of the A-, B-, and C-layers using
the drift times from the MDT and PDT systems. 
In each layer, all possible pairs of wire hits form the first segments.
They are then  iteratively  merged using a fit to a straight line.
The $\chi^2$ of the fit procedure is used to measure the segment
quality.  Segments in the B- and C-layers
are combined if they are consistent with being on a single straight line.
In the following, we denote as a BC-segment either a single B- or  C-segment, or the combination of a  B- and a C-segment. 
The final 2D segments require at least two wire
hits in different planes.
If a pair of wire hits in different planes of a given layer
is part of a segment 
composed of three or more planes, then only the segment with the highest hit multiplicity is retained.

After 2D segments are found, scintillator hits are associated with
each segment if they agree with the $\eta$ position for the forward
system and both the $\eta$ position and $\phi$ position
for the central system. At $|\eta|$ near 1.0, a
single particle
can go through both the PDT and MDT systems and segments made from
each system are combined if they are consistent with a single track.
The typical segment position resolution perpendicular to the wires is 0.7~mm while the segment angular
resolution is around 10~mrad for A-segments and 0.5~mrad for  BC-segments. 

A single muon can produce multiple segments,  for example, due to left-right ambiguities of the drift time.
To reduce combinatorics, only the three highest-quality segments arising from a given cluster of hits
are kept for the A-layer and only the two highest-quality segments for the BC-layer. The remaining ambiguities are resolved by the
local track finding.

\begin{figure}[!ht]
\begin{center}
\includegraphics[width=0.48\textwidth]{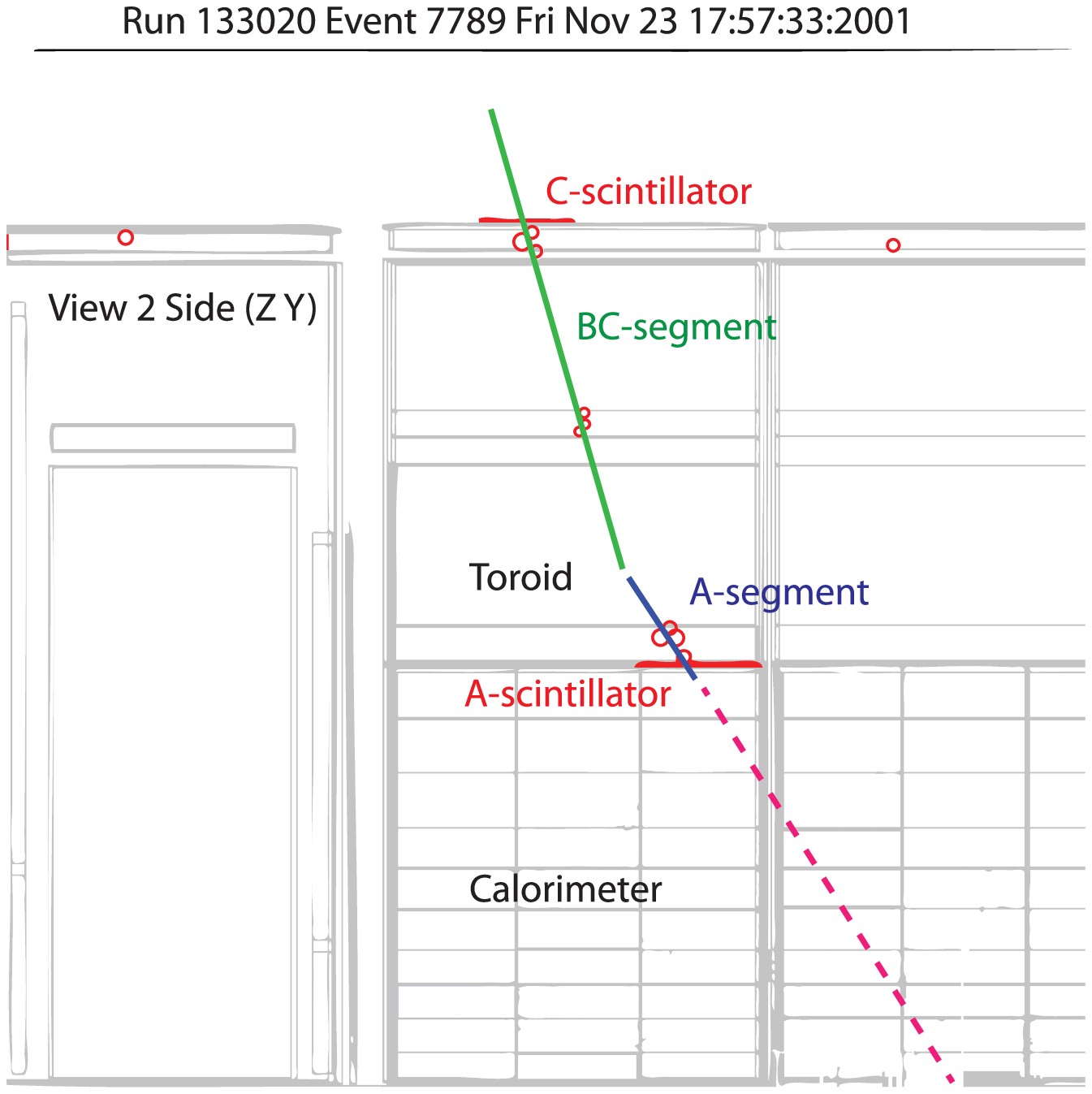}
\end{center}
\caption{Event display showing the A-segment and the BC-segment reconstructed from A-, B-, and C-layer hits.}
\label{fig:event_display}
\end{figure}

Figure~\ref{fig:event_display} shows an event display illustrating how segments are reconstructed using hits from the wire chambers and the scintillators.

\subsection{Local muon track finding}

Segments in the A-layer before the toroid and BC-layer after the
toroid are paired if they are consistent with the passage of a single
particle through the toroid. 
Non-paired A- and BC-segments are retained  in the list of local  muons
because muons can still be identified if the segments are matched to central tracks using the algorithm described in the next section.

\begin{figure}
\begin{center}
\includegraphics[width=0.48\textwidth]{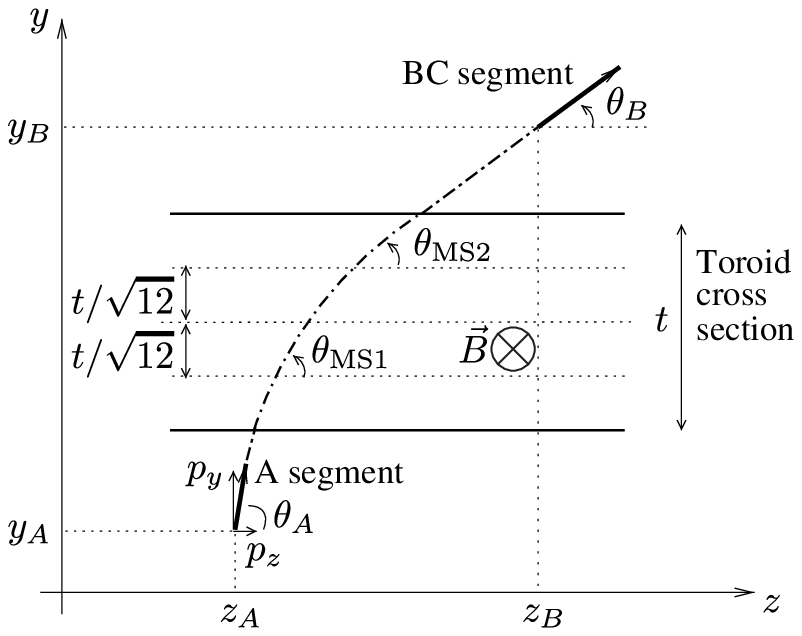}
\end{center}
\caption{Scheme of the local track fitting method. In this example, a muon is crossing the central top part of the muon system. Thus
the coordinate measured by the segment is $z$, the deviation plane is $(z,y)$, and the magnetic field ($\vec B$) is oriented along the $x$-axis.
See text for details.}
\label{local_track}
\end{figure}

For each compatible pair of segments, a muon track is reconstructed from
a fit  taking into account the magnetic field strength,
energy loss, and multiple Coulomb scattering in the toroids.
Figure~\ref{local_track} illustrates
the method and shows the different parameters and terms involved in the fitting procedure.
The covariance matrix for the random displacement in position and direction of a track due to multiple scattering
in a continuous material can be accurately reproduced
by just two independent random changes of direction, $\theta_{MS1}$ and $\theta_{MS2}$, if they occur
at the location $t (\frac{1}{2} + \frac{1}{\sqrt{12}})$ and $t (\frac{1}{2} - \frac{1}{\sqrt{12}})$
along the track, where $t$ represents the thickness of the material traversed~\cite{Deliot:2002ma}.
Using this property, the  five fitted parameters  defined in the deviation plane are therefore
the two scattering angles, $\theta_{MS1}$ and $\theta_{MS2}$;
the track position coordinate at layer A along the direction contained in the wire chamber plane and perpendicular to the wires,
$z_A$ in the example of Fig.~\ref{local_track}; 
the track angle at layer A, $\theta_A$; 
and the curvature $q/p_{\rm dev}$,
where $q$ is the muon charge and $p_{\rm dev}$ is the muon momentum at layer A projected to the deviation plane
($p_{\rm dev}=\sqrt{p^2_x+p^2_z}$ in the example of Fig.~\ref{local_track}).

The fit parameters define
the propagation of the muon track from the A-layer to the BC-layer by sequential small displacements, accounting for the deviation due to the magnetic field,
energy loss, and the deviation induced by the multiple scattering if crossing
one of the multiple scattering planes.
The fit is performed by a least squares method, where the  $\chi^2$ is constructed from four terms comparing the positions and angles of the track at the A- and BC-layers
to the segments' parameters, and two terms comparing the two scattering angles to their expected null value.

Once the fit has converged, the 3D track parameters are determined from the $\phi$ coordinate of the segments. 
To obtain the  momentum ($p=|\vec p|$) at the $p \bar p$ hard scattering vertex, an estimate of the energy loss in the calorimeter,
depending on the particle momentum and rapidity,  is added to the value determined at the A-layer.
The resolution of the momentum determined
solely by the muon system depends on $\eta$ and on whether the track has both B-
and C-layer hits.
This resolution can be measured by computing the difference between the local muon momentum and the matched central track momentum.
The results found for different types of local tracks and regions are given in Table~\ref{tab:muon_resolution}.
Primarily due to multiple scattering in the iron toroid,
these resolutions are significantly
worse than those obtained by the central tracking system, which
is about
$\frac{\sigma( p)}{p}=0.021\oplus 0.0025\cdot{ p_T}/{\mathrm{GeV}}$ at $\eta = 0$
(see Sec.~\ref{sec:momentum_resolution}).

\begin{table*}[!]\center
\begin{tabular}{cccc}
\hline\hline 
Region               &  Muon type & Parametrization for $\frac{\sigma (p)}{p}$ ($p$ and $\pt$ in GeV)  \\ \hline
 $|\eta|<0.88$       &   ABC      &   $\frac{p-2}{p}[0.21\oplus 0.011(p_T-2)]$ \\
 $|\eta|<0.88$       &   AB or AC &  $\frac{p-1.3}{p}[0.21\oplus0.016(p_T-1.3)]$ \\
 $0.88<|\eta|<1.1$   &   ABC      &  $\frac{p-2}{p}[0.28\oplus0.006(p-2)]$   \\
 $0.88<|\eta|<1.1$   &   AB or AC &  $\frac{p-2}{p}[0.30\oplus0.009(p-2)]$  \\
  $1.1<|\eta|<1.7$   &   ABC      &  $\frac{p-3}{p}[0.215\oplus0.007(p-3)]$\\
  $1.1<|\eta|<1.7$   &   AB or AC &  $\frac{p-3}{p}[0.28\oplus0.029(p-3)]$\\
 $1.7<|\eta|<2.0$   &   ABC &  $\frac{p-2.5}{p}[0.24\oplus0.005(p-2.5)]$ \\
\hline\hline
\end{tabular}
\caption{\label{tab:muon_resolution}
Parametrization for the local muon track  momentum resolution for different categories of muons.
}
\end{table*}

For a high-\pt\ muon within the acceptance of the muon system, the efficiency of the local
track reconstruction is around 85\%, which includes the small fraction ($<5\%$) of non-converged fits.
As noted previously,
several segments are kept for a given group of hits. The $\chi^2$ of the local track fit as well as the number of hits in the segments are used to choose
which pair of  segments is the best solution.

\subsection{Matching with central tracking}

\def \vb {\vec{B} }
\def \vpi {\vec{p}\,^{\rm{in}}  }
\def \vpo {\vec{p}\,^{\rm{out}} }
\def \vxi {\vec{x}\,^{\rm{in}}  }
\def \vxo {\vec{x}\,^{\rm{out}} } 
\def \vv {\vec{v} }
\def \vn {\vec{n} }
\def \vz {\vec{z} }
\def \ob {\vec{\Omega}^B}
\def \doms {\delta \vec{\Omega}^{MS}}
\def \el {dE_{\rm{loss}}}

To improve the resolution of the muon momentum,
local tracks are matched to  the precisely measured tracks of 
the central tracking system.
The matching is performed by propagating the tracks through the calorimeter,
taking into account the inhomogeneous magnetic field,  energy loss, and  multiple scattering.
The alignment between the muon system and the central tracking system
is better than approximately 0.5~mm. This is much smaller than
the impact  of multiple scattering (typically larger than 1~cm) and hence does not affect the matching.

We consider a relativistic particle of charge $q$, mass $m$ with initial position and momentum
$(\vxi,\vpi)$.
After traveling an infinitesimal distance $ds$ in a material, its position and momentum become
$(\vxo,\vpo)$, which can be written:
\begin{eqnarray} 
\vxo & = & \vxi +  \frac {\vpi }{| \vpi |} ds,  \\
\vpo & = & \vpi \left( 1 - \frac{\el}{| \vpi |} \right)
\nonumber \\ &&
+ \ob \times \vpi 
        + \doms \times \vpi  \label{vp} .  
\end{eqnarray}
In these equations,
$\el$ represents the   random energy loss in the calorimeter. Once integrated over the calorimeter thickness, the mean value of energy loss is around 3~GeV and its RMS is about 0.4~GeV.
The term:
\begin{eqnarray} 
\ob & = & - 0.3 \frac{q}{| \vpi |} \vb \;ds \label{eq:cm_vxo}, 
\end{eqnarray}
accounts for the curvature due to the magnetic field
(where the unit for the coefficient 0.3 is \gev$\cdot$T$^{-1}$$\cdot$m$^{-1}$),
and the term:
\begin{eqnarray} 
\doms & = & \frac{\delta \beta}{|\vv|} \left(\cos\alpha\, \vz 
        + \sin\alpha\, \vv \right) ,
\end{eqnarray}
represents the effects of the multiple scattering, where
$\delta \beta$ denotes the infinitesimal deviation from the momentum direction, $\vz$ is an arbitrary direction 
different from the momentum direction,  $\vv$ is defined as $\vv = \vz \times \frac{\vpi}{| \vpi |}$, and 
$\alpha$ is a random angle defining the direction of the deviation.
In a similar manner, it is possible to write the propagation of the track parameter error matrix~\cite{tuchming:HDR2010}.

This set of equations allows the propagation (with their respective error matrices) of either the local tracks inward to the center of the \dzero\ detector
or the central tracks outward to the muon system. 
For local tracks,
the inward propagation has precedence over the outward propagation.
For non-paired  A-segments, non-paired BC-segments, and  local tracks without a converged local fit, only the outward propagation is sensible since there is no accurate measurement of the local muon momentum.
In all cases, the quality of the matching between a local muon of parameters $P_m$ and error
matrix $V_m$  and a central track $P_c$ with error matrix $V_c$
is evaluated by a $\chi^2$ test: 
\begin{equation}
\chi^2 =  \left(P_m - P_c \right)^T\cdot \left(V_m+V_c \right)^{-1} \cdot
        \left(P_m - P_c \right). 
\end{equation} 
The mimimum  $\chi^2$ is used to choose which central track is matched to each of the local muons.

\subsection{Momentum measurement}
Since the local muon momentum resolution is inferior to
the resolution from the central tracking system,
the momentum of a muon matched to a central track is taken
to be the momentum measured in the central tracker.
Non-paired A- and BC-layer segments matched to central tracks also define muon objects, the momenta of which is taken from the matched tracks.

For muon candidates matched to central track without SMT hit,
the additional constraint that the track arises from the beam axis located at $(x_b, y_b)$ is used\footnote{
The transverse beam position is determined by averaging the primary vertex position over multiple beam crossings.
The beam is  up to two millimeters away from $(0,0)$ in the detector coordinate system.}.
This constraint yields a correction for the track curvature $\kappa$, which is given to first order by
\begin{equation}
\kappa \rightarrow  \kappa + (x_{b}\sin\phi_0 - y_{b}\cos\phi_0-d_0) \frac{V_{\kappa d}}{V_{dd}},
\end{equation}
where  $d_0$ is the distance of closest approach to the central axis $(x,y)=(0,0)$, $\phi_0$ is the track azimuthal angle at the point of closest approach,
and 
${V_{\kappa d}}$ and ${V_{dd}}$
are elements of the covariance matrix resulting from the central track reconstruction fit.
${V_{\kappa d}}$ is the covariance between $\kappa$ and $d_0$, while ${V_{dd}}$ is the squared uncertainty on $d_0$.
The correction is propagated to the track transverse momentum which is proportional to $\frac 1 \kappa$.
Central tracks without hits in the SMT have a relative resolution on transverse momentum $\frac {\sigma(\pt)}{\pt}=\frac {\sigma(\kappa)}{\kappa}$
of typically 25\% for $\pt=45$~GeV.
The correction improves the relative resolution to 15\%--18\%, compared to the typical resolution of 10\% for tracks with SMT hits.

\section{Muon identification}
\label{sec:identification}
\makeatletter{}
\def\loose{loose}
\def\medium{medium}
\def\mediumnseg3{tight}

        \def\trackloose{loose}
\def\trackmedium{tight}
\def\tracknewmedium{medium}
\def\tracktight{mediumSMT}

\def\TopScaledLoose{scaledLoose}   
\def\TopScaledMedium{scaledMedium}
\def\TopScaledTight{scaledTight}
\def\NPTight{tight}          
\def\TrkTight{trkTight}         
\def\TrkLooseScaled{trkScaledLoose}
\def\TrkTightScaled{trkScaledTight}
\def\deltaR{jetIso}

A muon candidate is primarily defined by
(i)~the presence of a  local muon in the muon system.
Additionally, for most physics analyses, the local muon has to
to be (ii)~matched to a track in the central tracker.
For high-\pt\ physics,
(iii)~the absence of significant activity around the muon trajectory,
both in the calorimeter and in the central tracker  may also be demanded.
For each of these three criteria, different identification quality categories are defined,
which are briefly presented in the following sections.
The efficiencies for the different categories and the experimental techniques used to measure them in data and simulated events are summarized in Sec.~\ref{sec:performance}.

Despite the relatively high amount of energy lost by a muon in  the calorimeter,
the energy deposit of muons in an individual cell is close to the threshold level of the calorimeter noise-suppression algorithm~\cite{Vlimant:2005ur},
and is therefore not well measured.
Thus, the calorimeter information is not exploited to identify high-\pt\ muons
but is used for muon identification in heavy flavor analyses.

\subsection{Identification criteria in the muon system}\label{ssec:muid}

For the identification of local muons, three categories, {\em \loose}, {\em \medium}, and {\em \mediumnseg3},  are defined as follows.

\begin{itemize}
\item
A local muon is {\em \loose} if (a)~it has at least one scintillator hit and at least two wire hits in the A layer 
of the muon system, or (b)~at least one scintillator hit and at least two wire hits in the BC layers.

\item
In the general case, a local muon is  {\em \medium} if it meets both conditions (a) and (b), except that for $|\eta|<1.6$, the BC scintillator requirement is dropped.
For the particular case of the bottom part of the detector where
the support structure for the calorimeter is located ($\frac{5\pi}{4}<\phi<\frac {7\pi} {4}$ and $|\eta|<1.6$), a local muon is {\em \medium} 
if it fulfills either condition~(a) or~(b) above.
In the particular case of a low-\pt\ muon, a local muon is {\em \medium} if it fulfills condition~(a) 
above and, additionally, its probability to reach the BC-layers is less than~70\% (due to energy loss in the toroid).
\item
A local muon is  {\em \mediumnseg3} if it belongs to the category of {\em \medium} muons that meet both conditions (a) and (b), except that for $|\eta|<1.6$, the BC scintillator requirement is dropped.

\end{itemize}
The number of  categories is doubled depending on whether or not a veto against cosmic muons is required.
The veto criterion demands that the scintillator hit times in each layer, if available, be consistent within 10~ns 
with those of a particle moving at the speed of light from the primary vertex. It has a typical efficiency of about 98.5\% for high-\pt\ muons.

\subsection{Identification criteria in the central tracker}\label{ssec:track}

For the identification of muon tracks in the central tracking system, four quality categories are defined:
 {\em \trackloose}, {\em \tracknewmedium}, {\em \tracktight}, and {\em \trackmedium}, as follows. 
\begin{itemize}
\item
A muon track containing SMT hits is defined as {\em \trackloose} if $|\dca|<0.04$~cm,
where \dca\ is the track distance of closest approach to the beam axis.
This requirement is changed to $|\dca|<0.2$~cm for tracks without SMT hits.
The {\em \trackloose}  quality
is a good choice for analyses
that do not need  the most accurate muon momentum measurement.

\item
A  muon track is {\em \trackmedium} if it fulfills the {\em \trackloose} requirements and if
$\chi^2/N_{\rm DOF}<4$,
where $\chi^2$ is the result of the fit
used for reconstruction of the track 
in the central tracking system and 
$N_{\rm DOF}$ is the  number of degrees of freedom.
This  $\chi^2$ requirement introduces notable efficiency losses 
because of wrongly assigned hits arising from tracks from  other $p\bar p$ interactions in the same bunch crossing
at
luminosity\footnote{ Over the course of Run~II, the distribution of luminosity ranges from 1 to $400\times 10^{30}$~cm$^{-2}$s$^{-1}$.
Its average value is around $140\times 10^{30}$~cm$^{-2}$s$^{-1}$ with a RMS of  $70\times 10^{30}$~cm$^{-2}$s$^{-1}$.
}
higher than typically $150\times 10^{30}$~cm$^{-2}$s$^{-1}$.
To overcome this issue, the {\em \tracknewmedium} quality has been defined using looser $\chi^2$ requirements.

\item
A  muon track is defined as {\em \tracknewmedium} if it is of {\em \trackloose} quality  and if, in addition,
$\chi^2/N_{\rm DOF}<9.5$.
Furthermore, at least two hits in the CFT
are required to
further discriminate against accidental combinations of hits in the SMT at high instantaneous luminosity.
\item
A muon track is {\em\tracktight} if it fulfills the {\em \tracknewmedium} requirements and, in addition, has hits in the SMT. This results in a lower rate from fake and misreconstructed tracks, and also better momentum resolution compared to the other categories (see Sec.~\ref{sec:track_efficiency}).
 
\end{itemize}

\subsection{Identification of isolated muons}\label{ssec:iso}

\begin{table*}[!]
\center
\begin{tabular}{lcccccc }
\hline\hline
Operating point     & \itrk          & \ical          & $\itrk/\pt^\mu$ & $\ical/\pt^\mu$   & $\dr({\rm \mu,jet})$\\
\hline
{\em \TopScaledLoose}      & --            & --            & $<\!0.20$  & $<\!0.20$ &   $>0.5$  \\
{\em \TopScaledMedium}     & --            & --            & $<\!0.15$  & $<\!0.15$ &  $>0.5$\\
{\em \TopScaledTight}      & --            & --            & $<\!0.10$  & $<\!0.10$ &   $>0.5$\\
{\em \NPTight}             & $<\!2.5\,\gev$ & $<\!\phantom{1}2.5\,\gev$& --      & --&  $>0.5$        \\
{\em \TrkTight}            & $<\!2.5\,\gev$ & $<\!10.0\,\gev$  & --      & --        &  $>0.5$\\
{\em \TrkLooseScaled}      & --            & --            & $<\!0.25$  & $<\!0.40$  &  $>0.5$\\
{\em \TrkTightScaled}      & --            & --            & $<\!0.12$  & $<\!0.40$  &  $>0.5$\\
{\em \deltaR}              & --            & --            & --        & --          &  $>0.5$\\
\hline\hline
\end{tabular}
\caption{\label{tab:isoworkingpoints}
Definitions of
isolation operating points. }

\end{table*}

We select isolated muons arising from the primary vertex
by rejecting secondary muons from semi-leptonic decays of $b$ or $c$ quarks,
which are surrounded by additional particles due to quark fragmentation and other heavy hadron decay products.
Three basic discriminating variables are formed as follows.

\begin{itemize}

\item 
$\dr({\rm \mu,jet})
=\sqrt{
 \Delta\eta ^2  ({\rm \mu,jet})
+ \Delta\phi ^2  ({\rm \mu,jet})
}$, is the closest distance in $(\eta,\phi)$ space  of the muon to any  jet with $\pt>15$~GeV, where the jets are 
reconstructed from energy deposits in the calorimeter using
an iterative midpoint cone algorithm~\cite{Blazey:2000qt} with a cone radius  ${\cal R}= 0.5$.

\item 
$\itrk \equiv \sum_{\{{\rm tracks}\in\dr<0.5\}} \pt^{\rm track}$ is the scalar 
sum of transverse momenta of all tracks inside a $\dr({\mu,\,\rm track}) < 0.5$ cone around the muon track with the 
exception of the muon track itself.
To reject the contributions of tracks arising from other $p\bar p$ interactions in the same bunch crossing,
the requirement  $\Delta z_0({\rm \mu, track})<2$~cm is demanded for each track in the sum,
where $z_0$ is the coordinate of the track at the point of closest approach to the beam axis\footnote{The  beam interaction region is distributed along the $z$-axis following approximately
a Gaussian function of width $\simeq$ 25--30~cm.}.

\item
$\ical \equiv \sum_{\{{\rm clusters}\in0.1<\dr<0.4\}} \et^{\rm cluster}$, is the scalar sum of transverse 
energies of all calorimeter clusters inside a hollow cone around the muon defined by $0.1<\dr({\mu,\,\rm cluster}) < 0.4$.
Only the energy deposits in the electromagnetic calorimeter and the first fine sampling layers of the hadron calorimeter 
are considered to reduce the  impact of noise and  other $p\bar p$ interactions in the same bunch crossing.
\end{itemize}

We also employ as isolation variable $\itrk/\pt^\mu$ and $\ical/\pt^\mu$ which 
offer higher efficiencies for high-$\pt$ muons and 
more stringent rejection against secondary leptons from $b$- and $c$-quark decays at low $\pt$.
Because of the $\Delta z_0({\rm \mu, track})<2$~cm requirement which  rejects tracks 
from  other $p\bar p$ interactions in the same bunch crossing efficiently, the quantities  $\itrk$ and $\itrk/\pt^{\mu}$ are more robust at high
instantaneous luminosity conditions, compared to  $\ical$ and $\ical/\pt^{\mu}$.

Based on these five variables, several isolation criteria are defined as shown in Table~\ref{tab:isoworkingpoints}.

\section{Performance}
\label{sec:performance}
\makeatletter{}
\begin{figure}[t]
\begin{center}
\includegraphics[width=0.48\textwidth]{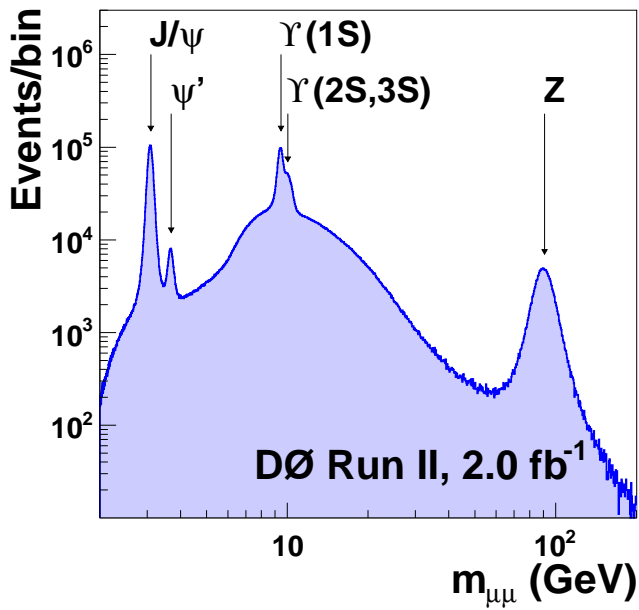}
\end{center}
\caption{[color online] The invariant mass spectrum for dimuon events reconstructed with the \dzero\ detector. The observed resonances are indicated by arrows.}
\label{fig:performance}
\end{figure}

The ability of the \dzero\ detector to identify and reconstruct muons is illustrated in Fig.~\ref{fig:performance},
where the dimuon invariant mass ($m_{\mu\mu}$) spectrum of events with two oppositely charged 
and isolated muon candidates is shown for a representative subset of data collected in Run~II of the Tevatron.
To obtain this spectrum, the selection requires two {\em \medium} muons matched to {\em \trackmedium} tracks with $p_T>3~$GeV.
Both muons have to be isolated according to  $\itrk,\ical<4~\gev$, and at least one of them also has to be tightly isolated according to  $\itrk,\ical<2.5~\gev$.
For this spectrum and in the following we impose data quality requirements.
Due to the $\pt>3$~\gev\ requirement, the lightest observed resonance is from the production of $\jpsi$ mesons. It is followed by resonances from the $\psi'$ and $\Upsilon$ mesons, as well as the $Z$ boson.

The performance of the muon reconstruction can be quantitatively assessed in terms of 
identification efficiency, fake rate, and momentum resolution. 
The methods to measure these performances for high-\pt\ muons and the results are discussed below.
More details can also be found in Refs.~\cite{tm2540,{tm2541}}.

\subsection{Tag-and-probe method}

To measure the identification and reconstruction efficiency for high-\pt\ muons in data, we apply the ``tag-and-probe'' technique based on a \zgmumu\  sample.
The efficiency is factorized into three
independently measured components: the muon efficiency, the track efficiency, and the isolation efficiency

In this method, one muon candidate with tight selection requirements serves as a tag, whereas the other candidate serves as a probe and is used for the efficiency measurement.
Additional requirements for consistency with the decay of a $Z$ boson are imposed on both muon candidates, e.g., a requirement on the invariant mass of the dimuon system.
To remove any residual bias due to trigger requirements,  only specific triggers are used to select \zgmumu\ candidate events. Each of the muon candidates in a given event can serve as a tag and as a probe.

The efficiency to identify and reconstruct a high-\pt\
muon exhibits a dependence on  instantaneous luminosity.
Trigger conditions prescaled at high luminosity modify the luminosity spectrum
and bias the efficiency measured as an average over the actual running conditions.
We therefore perform a reweighting of the $\zgmumu$ sample to bring its  luminosity spectrum into agreement with that found in a subset of events selected with the requirement of  having at least one muon candidate. This subset is typically used in physics analyses in final states with one or more muons.

\subsection{Efficiency of muon system reconstruction}
\label{sec:muon_efficiency}

\begin{figure*}
\begin{center}
\subfigure[]
{\includegraphics[width=0.32\textwidth]{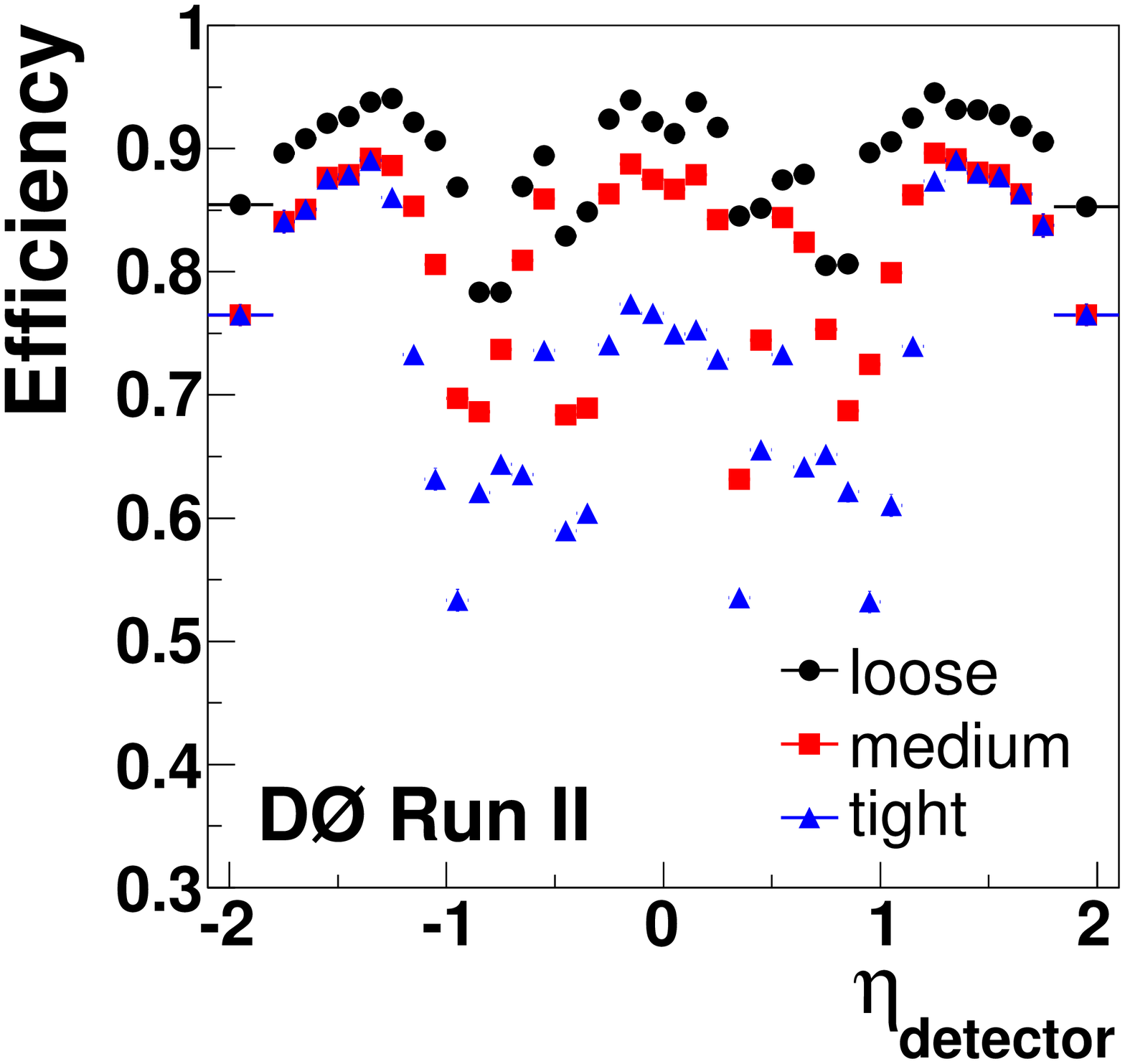}}
\subfigure[]
{\includegraphics[width=0.32\textwidth]{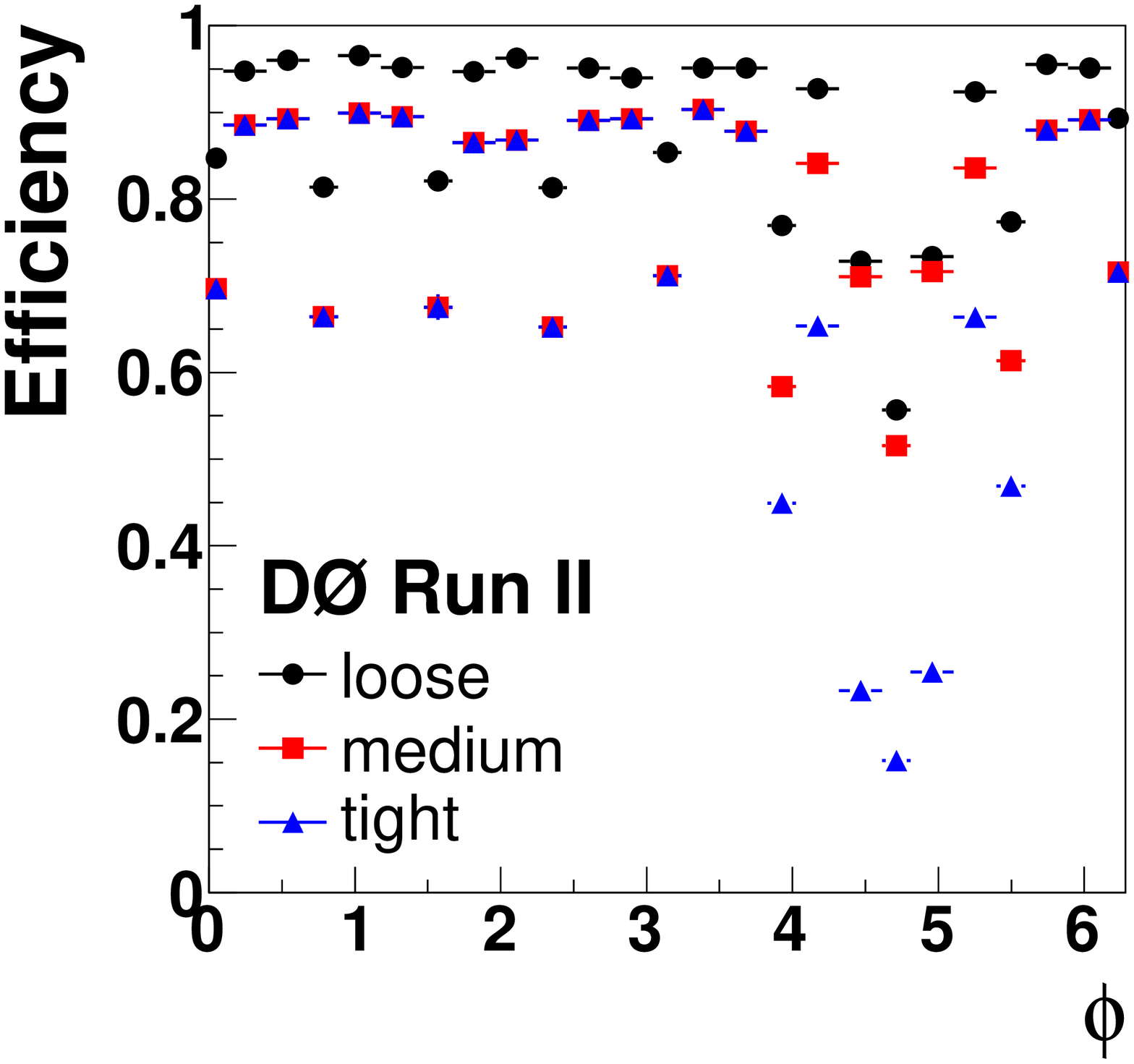}}
\end{center}
\caption{[color online] Efficiencies of the identification criteria ({\em \loose}, {\em \medium}, and {\em \mediumnseg3}) in the muon system
as functions of (a) $\etadet$ and (b) $\phi$.
}
\label{fig:muon_efficiency}
\end{figure*}

The tag-and-probe selection for the measurement of the efficiency of the D0 muon system to identify and reconstruct muons is summarized as follows.
The {\em tag} object is required to be a local {\em \medium} muon   matched to a central track of {\em \trackmedium} quality with $p_T>30~$GeV and isolated with $\itrk<3.5~$GeV and $\ical<2.5~$GeV.
It must have  fired a single muon trigger and the absolute value of the
A or B layer scintillator time has to  be less than~7~ns.
The {\em probe}  is required to be a central track of quality {\em \trackmedium} with $p_T>20~$GeV and isolated with $\itrk<3.5~$GeV and $\ical<2.5~$GeV, matched within $\dr<0.5$ to the local muon track.
The  {\em tag} and the {\em  probe} tracks have to be of opposite electric charge,
and have to fulfill  $|\Delta z_0|<2$~cm and  $\dr>2$.
Cosmic rays are suppressed by demanding $\pi -|\phi_{\rm tag}-\phi_{\rm probe}| + |\pi - \theta_{\rm tag} - \theta_{\rm probe}| > 0.05$.

The measured efficiencies for  {\em \loose}, {\em \medium}, and {\em \mediumnseg3} muons 
 are shown in Fig.~\ref{fig:muon_efficiency} as  functions of
$\etadet$ and $\phi$, where $\etadet$ is the angular coordinate (with respect to the center of the D0 detector)
of the position where the muon trajectory traverses the A-layer of the muon system.
Note that because of the spread in $z$ of the primary vertex distribution,  $\etadet$ can be different from the actual pseudorapidity of the muon.
In the figure, 
the bin-to-bin rapid variations of the measured efficiencies are due to the small non-instrumented part of the detector between the wire chambers, and, to a lesser extent, between the scintillators.
Because of the support structure of the calorimeter,   the bottom part of the muon system is less instrumented which is reflected in lower efficiency  for $\frac{5\pi}{4}<\phi<\frac {7\pi} {4}$, $|\etadet|<1.6$.
The average reconstruction efficiencies are 88.9\%, 80.8\%, and 72.0\% for the {\em \loose},  {\em \medium}, and {\em \mediumnseg3} operating points, respectively.
If the cosmic veto is not required, these efficiencies increase to 90.9\%, 82.5\%, and 73.1\%, respectively.
The statistical uncertainty on these efficiencies is of the order of 0.1\% and thus negligible.
We consider various sources of systematic uncertainty that may bias the measurement of these  efficiencies.
The dominant source of relative systematic uncertainty is  possible background contamination (0.8\%--1.1\%), with smaller contributions from $p_T^Z$ dependence (0.3\%), from the tag-and-probe technique (0.2\%), and from pattern recognition~(0\%--0.4\%). The relative uncertainties amount in total to  0.9\%--1.2\%.

There are only minor changes in the identification efficiency of the muon system over the course of Run~II. A small fraction of this effect is due to a slight dependence on instantaneous luminosity, while most of it is driven by the number of operating PDTs. 
The efficiency of single hit reconstruction and assignment to tracks  is dominated by alignment effects. Variations over the course of Run~II are small.

\subsection{Efficiency of muon central track reconstruction}
\label{sec:track_efficiency}
\begin{figure*}
\begin{center}
\subfigure[]{\includegraphics[width=0.32\textwidth]{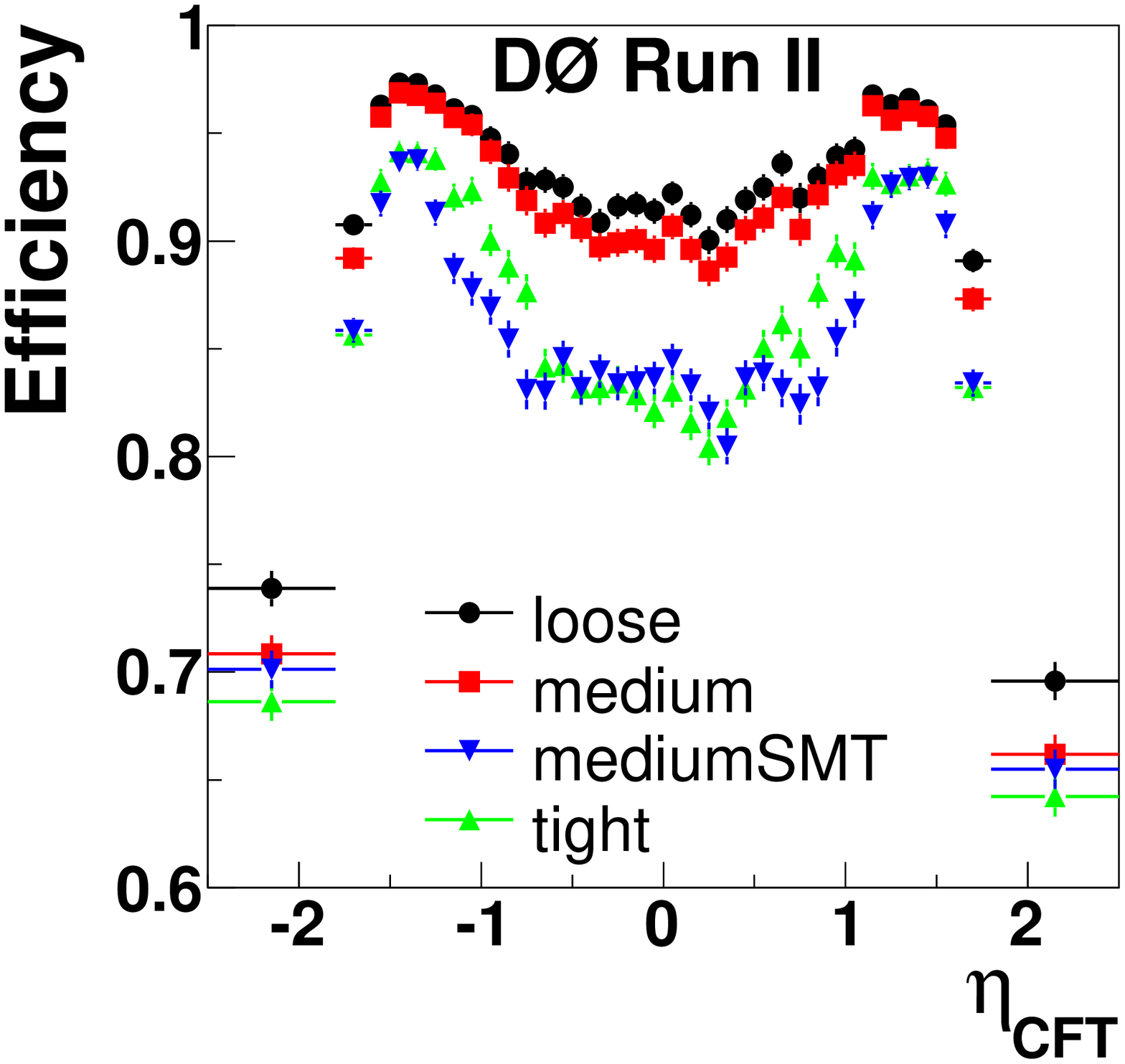}}
\subfigure[]{\includegraphics[width=0.32\textwidth]{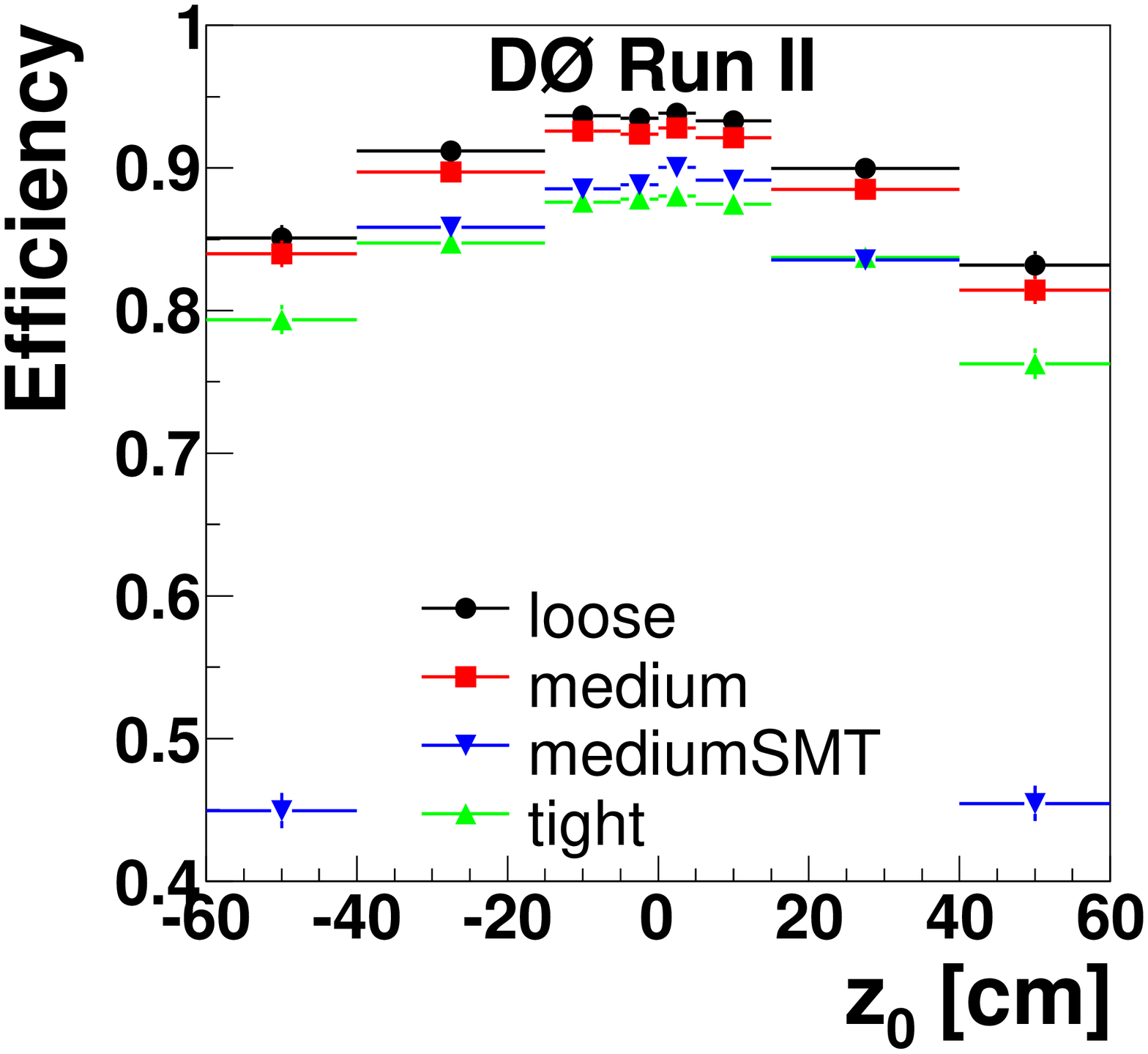}}
\subfigure[]{\includegraphics[width=0.32\textwidth]{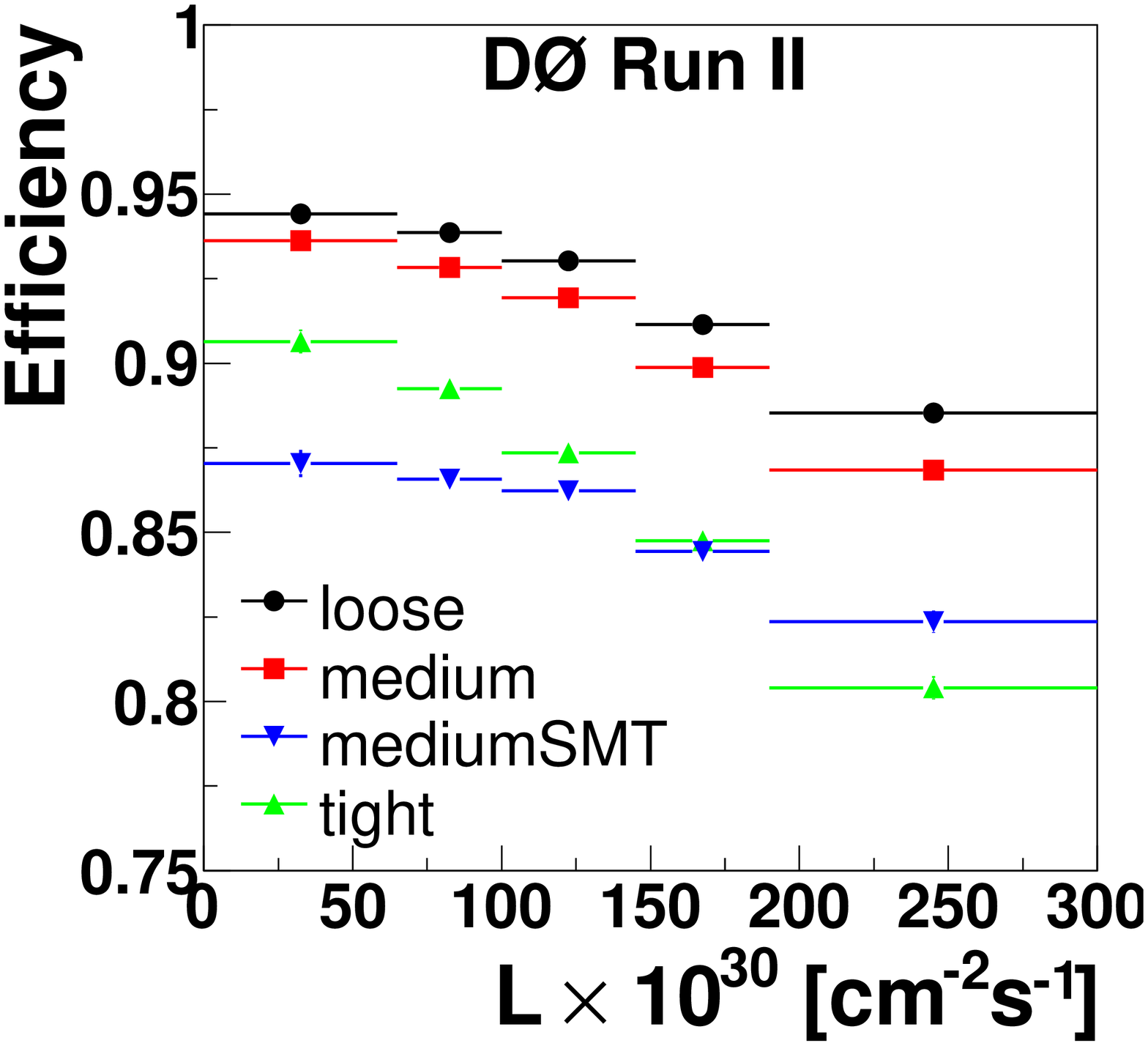}}
\end{center}
\caption{ [color online] Efficiencies of the identification criteria ({\em \trackloose}, {\em \tracknewmedium}, {\em \tracktight}, and {\em \trackmedium}) in the tracking system
as  functions of (a) $\etacft$, (b) $z_0$, and (c) instantaneous luminosity (\lumi).}
\label{fig:track_efficiency}
\end{figure*}

The tag-and-probe selection for the measurement of the efficiency of the \dzero\ central tracker to reconstruct muon tracks is summarized as follows.
The {\em tag} object is required to be a local {\em \loose} muon
matched to a central track of {\em \trackmedium}  quality with $p_T>30~$GeV and $|\dca|<0.2~$mm,  and isolated with $\itrk<3.5~$GeV and $\ical<2.5~$GeV.
The {\em probe} is a local muon track of {\em \loose} quality, with $p_T>15~$GeV as measured in the muon system, and isolated with $\ical<2.5~$GeV.
The {\em tag} and the  {\em probe} must fulfill $\dr>2$, and their respective scintillator hit times at  either the A- or B-layer have to match within a 6~ns window.
For these events, the trigger requirement consists of a dimuon trigger with no explicit central track condition.

The measured efficiencies are shown in Fig.~\ref{fig:track_efficiency}, as a function
of $\etacft$, $z_0$ and luminosity, where  $\etacft$ is the angular coordinate
of the outermost intercept between the muon trajectory and the CFT detector volume.
The average efficiencies are 91.6\%,  90.5\%, 84.6\%, and  86.2\%, for the {\em \trackloose},
{\em \tracknewmedium}, {\em \tracktight}, and {\em \trackmedium} operating points, respectively.
Various sources of relative systematic uncertainty that may bias our efficiency measurement are considered:
modeling of the primary vertex position along the beamline~(0.7\%--0.8\%),  possible background contamination~(0.5\%),
the tag-and-probe technique~(0.3\%), and the jet multiplicity in the final state~(0.1\%--1.4\%). The relative statistical uncertainty is of the order of 0.1\% and thus negligible with respect to the total systematic uncertainty.
The relative uncertainties amount in total  to 1.1\%--1.6\%.

There were changes, some of them substantial, in the reconstruction efficiency of muon tracks in the central tracker over the course of Run~II.
The most significant changes,
resulting in an increase of the track reconstruction efficiency of the order of 1\% despite detector aging effects,
occurred in 2006 after 1~\fbi\ of data had been collected,
when an additional Layer~0 was installed in the SMT~\cite{Angstadt:2009ie}.
A similar increase occurred in 2009 after  6~\fbi\ of data had been collected, when a large fraction of the non-responsive modules of the SMT were recovered.
The pronounced dependence of the track reconstruction efficiency on  instantaneous  luminosity resulted in an increasingly adverse effect in the second half of Run~II.
As the detector aged,
the reconstruction efficiency decreased in the central region of the CFT and, to a smaller degree,
in the inner layers of the SMT towards the end of Run~II.
The former effect is attributed to reduced scintillation light yield resulting from radiation damage,
which has the largest impact on reconstruction efficiency at central pseudorapidity due to the smaller
effective path length of the muons through individual fibers at those pseudorapidities.
The reduction in reconstruction efficiency in the inner layers of the SMT is attributed to radiation damage to the silicon sensors.
As a result, their operating high voltage had to be increased in order to fully deplete the sensor. However,
this voltage is limited to 150~V for the inner SMT barrel layers (except Layer 0),
thus resulting in a reduced active sensitive region for some of the inner sensors towards the end of data taking.
The innermost Layer 0 remained fully depleted throughout Run~II.

\subsection{Efficiency of isolation requirements}
\label{sec:iso_efficiency}
\begin{figure*}
\begin{center}
\subfigure[]
{\includegraphics[width=0.32\textwidth]{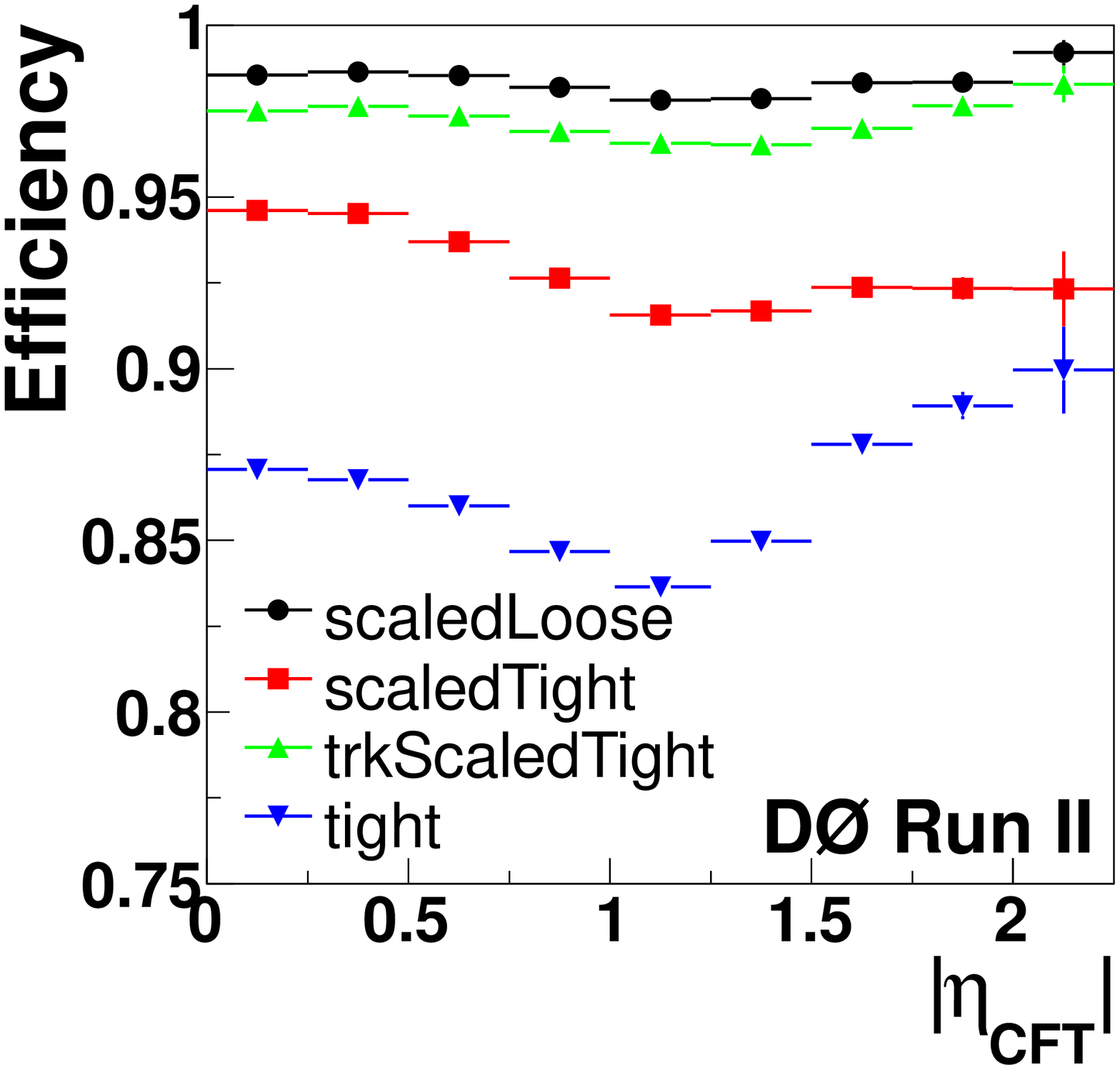}}
\subfigure[]
{\includegraphics[width=0.32\textwidth]{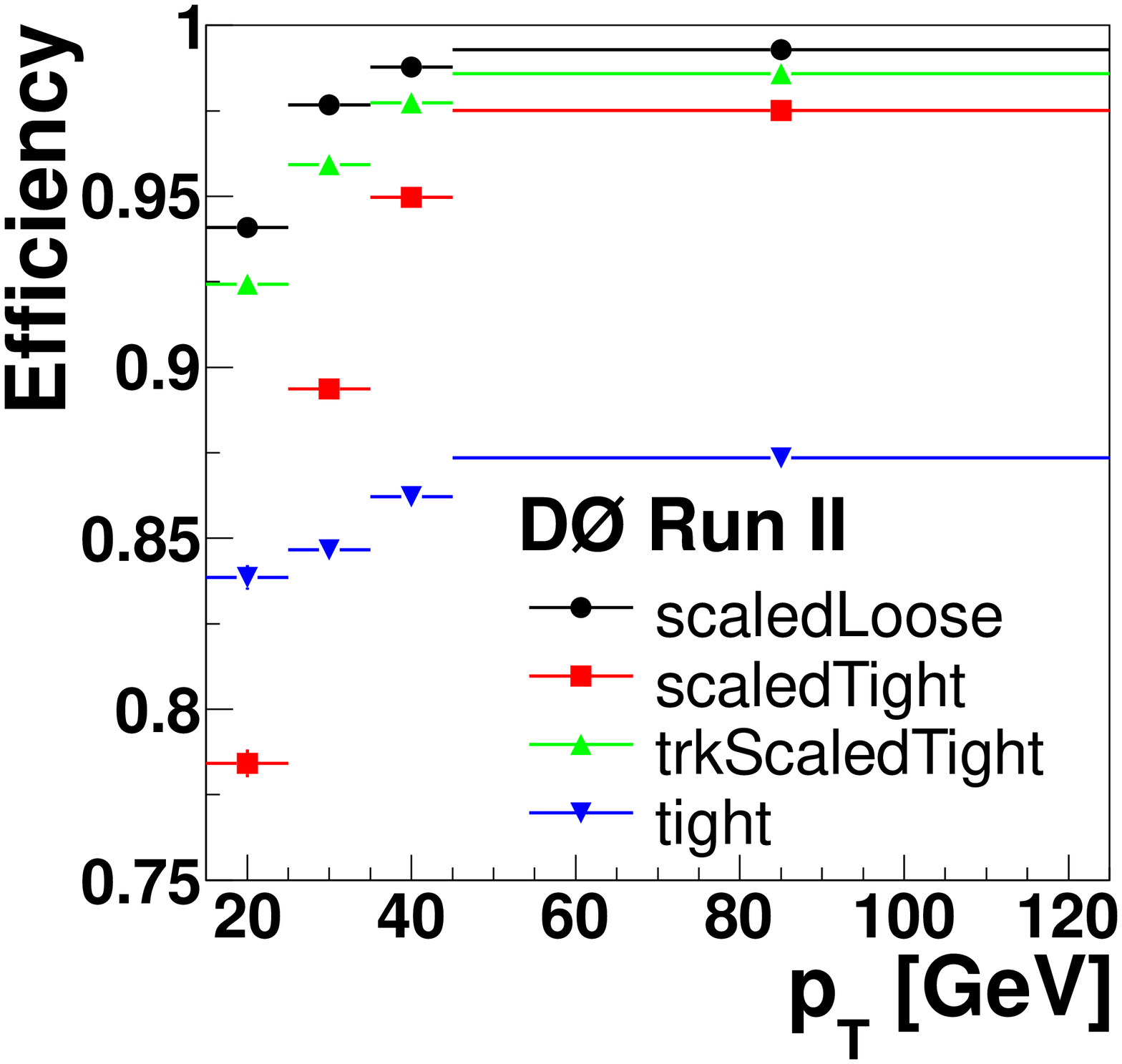}}
\subfigure[]
{\includegraphics[width=0.32\textwidth]{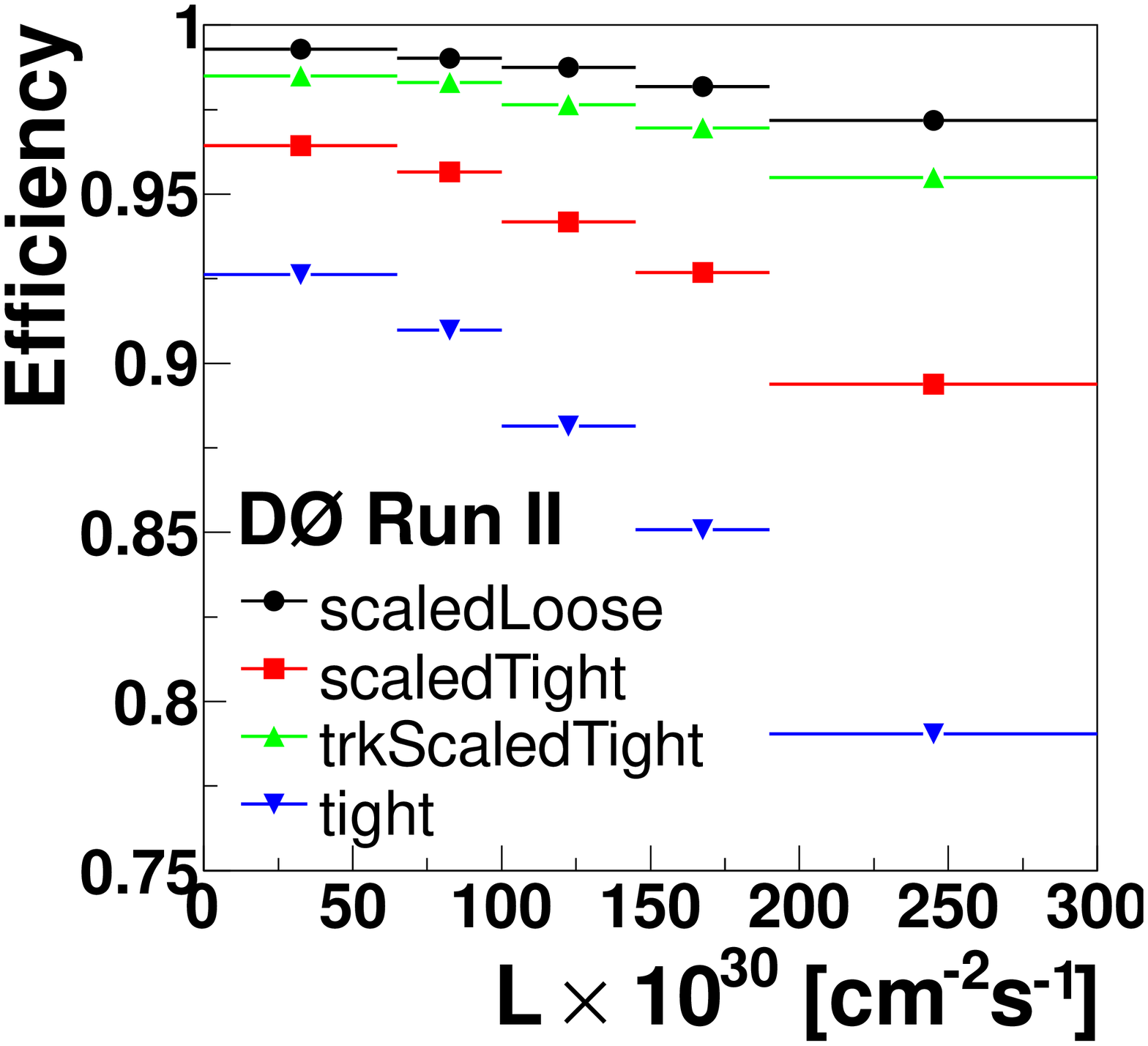}}
\end{center}
\caption{ [color online] Efficiencies of the isolation criteria
as  functions of (a) $\etacft$, (b) $\pt$, and (c) instantaneous luminosity (\lumi).}
\label{fig:iso_efficiency}
\end{figure*}

The tag-and-probe selection for the measurement of the isolation efficiency is summarized as follows.
Both the {\em tag} and {\em probe} objects are required to be  local muons of  {\em \loose} quality, with  $p_T>8~$GeV,
matched to  central tracks of {\em \trackloose} quality, with $p_T>15~$GeV and $|\dca|<0.4~(2)~$mm if  matched (not matched) to hits in the SMT.
In addition, the tag muon has to be isolated according to  $\itrk<2.5~$GeV and $\ical<10~$GeV, and a dimuon trigger with no explicit isolation requirement has to have fired.
The  {\em tag} and the {\em  probe} tracks have to be of opposite charge  and must fulfill  $|\Delta z_0|<2$~cm,  $\dr>2$,
 $\pi -|\phi_{\rm tag}-\phi_{\rm probe}| + |\pi - \theta_{\rm tag} - \theta_{\rm probe}| > 0.05$, and $70~\GeV<m_{\mu\mu}<120~\GeV$.

A sample of measured efficiencies for the operating points defined in Table~\ref{tab:isoworkingpoints} is shown in Fig.~\ref{fig:iso_efficiency} as functions of  $\etacft$, $\pt$,
and luminosity. All these efficiencies are measured with respect to the
{\em \deltaR}  criterion ($\dr({\rm \mu,jet})<0.5$)
to factorize out the dependence on event topology.
The {\em \deltaR} criterion has an efficiency of 95.8\% for the particular topology of
inclusive \zgmumu\  selection.

The average efficiencies relative to the  {\em \deltaR} criterion are reported
in Table~\ref{tab:iso_effi}. They are  in the range  87.3\% to 98.6\%.
\begin{table}[!ht]
\center
\begin{tabular}{lc }
\hline\hline
Operating point     & Efficiency (\%) relative \\
&to $\dr({\rm \mu,jet})<0.5$ \hfill \\
\hline
{\em \TopScaledLoose}      & 98.4\\
{\em \TopScaledMedium}     &  97.3\\
{\em \TopScaledTight}      &  93.6\\
{\em \NPTight}             &  87.3\\
{\em \TrkTight}            &  94.1\\
{\em \TrkLooseScaled}      &  98.6\\
{\em \TrkTightScaled}      &  97.3\\ 
\hline\hline
\end{tabular}
\caption{\label{tab:iso_effi}
Average efficiencies for the different
isolation criteria measured with respect to the {\em \deltaR} criterion. }
\end{table}
For each criteria, the relative statistical uncertainty is of the order of 0.1\% and thus negligible.
We consider various sources of relative systematic uncertainty that may bias our measurement of the isolation efficiency: potential tag-and-probe biases (0.3\%), possible background contamination (0.2\%), correlation to the identification of muons in the muon system~(0.1\%) and to the reconstruction of tracks in the central tracker~(0.1\%), as well as the mismodeling of the luminosity spectrum~(0.1\%--0.8\%).
The total relative uncertainty amounts to 0.4\% for the efficiency of the  {\em \deltaR} criterion.
For the efficiencies of the other operating points with respect to  the  {\em \deltaR} criterion, the total relative uncertainties
are 0.4\%--0.8\%.

There have been variations in the isolation efficiency of muon candidates during Run~II, up to several percent for some working points.
They are due to the dependence of the isolation efficiency on the detector occupancy,
which depends on instantaneous luminosity.
This effect is mostly driven by the calorimeter isolation, for which it is not  possible to control the effect of pile-up from  other $p\bar p$ interactions in the same bunch crossing. The {\em \deltaR} efficiency is almost independent ($<1\%$ variation) of luminosity.
The isolation conditions based on the tracking system  are less affected by the presence of  other $p\bar p$ interactions in the same bunch crossing as they benefit from the requirement  $\Delta z_0({\rm \mu, track})<2$~cm imposed on the tracks considered in the isolation calculation.
This is one of the reasons why working points using mostly track isolation ({\em \TrkTight}, {\em \TrkLooseScaled}, and {\em \TrkTightScaled}) were defined.
The {\em \NPTight} working point shows the most pronounced dependence on instantaneous luminosity.

\subsection{Overall efficiency for high-\pt\ muons}
The overall  high-\pt\ muon reconstruction and identification efficiency is obtained by convoluting
the efficiency maps of the local muon, tracking, and isolation criteria with the
distribution in the parameter space of ($\pt,\eta,\phi,z_0,L$).
Once acceptance requirements have been defined, typically $\pt>20~\gev$ and $|\eta|<2$, the correlation between the different efficiency maps are weak.
An approximate of this convolution at the $10\%$ level can therefore be obtained from the product of average efficiencies quoted in the previous sections.
For example, for a  $p\bar p\to W\to\mu\nu$ event, the efficiency for the acceptance cuts $\pt>20~\gev$, $|\eta|<2$, and $|z_0|<40$~cm is $\sim 64\%$.
Because of the background level, strict selection requirements are generally employed,
such as {\em \medium} local muon, {\em \tracktight} track, and {\em \NPTight} isolation criteria.
Their overall efficiency is $\sim 80.8\%\times 84.6\%\times 87.3\%\times 95.8\% \sim 57\%$.

\subsection{Momentum resolution}
\label{sec:momentum_resolution}

The resolution of the muon momentum measured in the tracking system can be modeled by:
\begin{equation}
\sigma \left (  \frac{1}  {\pt} \right ) = 
\frac {R_{\CFT}^2} {\LARM^2} \left ( A \oplus \frac{B\sqrt{\cosh \eta}}{\pt} \right ),
\label{eq:resol2}
\end{equation}
where
$A$ is the resolution term related to the detector alignment and hit resolution, $B$  describes the effect of multiple Coulomb scattering,
$R_{\CFT}=52$~cm is the outer radius of the CFT detector,
and $\LARM$ is the radius corresponding to the outermost CFT hit along the track.
The term $R_{\CFT}/\LARM$ is a correction that accounts for the lever arm used to measure the track momentum;
this ratio is usually unity for tracks within the full acceptance of the CFT ($|\etacft|<1.6$).

The resolution parameters $A$ and $B$ can be measured
by comparing samples of $\zmumu$ and $\jpsimumu$ data 
to Monte Carlo (MC) events generated by the  \textsc{alpgen}~\cite{Mangano:2002ea} and \textsc{pythia} generators~\cite{Sjostrand:2006za}, respectively.
We use these two resonances, which correspond to two different scales of momentum, to disentangle the roles of $A$ and $B$ in the resolution.
For both data and MC, the selections of the samples demand two isolated, acollinear, oppositely charged muons, either  with $\pt > 3$~GeV to obtain the $\jpsimumu$ dominated sample, 
or with $\pt > 20$~GeV  to obtain the  $\zgmumu$ dominated sample.
The opposite sign requirement of the data $\jpsimumu$  selection is reversed to obtain a same-sign $\mu^\pm\mu^\pm$ control sample that models the background 
under the $\jpsimumu$ invariant mass peak.

In the MC samples, we apply a random smearing of the generated true muon momentum 
inspired by the form of the resolution Eq.~(\ref{eq:resol2}).
The random smearing is applied according to:
\begin{equation}
\frac{q}{p_T} \rightarrow (1 + S) \frac{q}{\pt} +  G   \frac{R_{\CFT}^2} {\LARM^2}  \left(A \oplus \frac{B \sqrt{\cosh \eta}}{\pt}\right),
\label{eq:resolution_smearing}
\end{equation}
where $q$ is the particle charge, the  $S$ parameter represents the potential difference in momentum scale between data and MC, and $G$ is a random number following a Gaussian distribution of mean zero and width one.
We compare the resulting smeared dimuon invariant mass spectra to their data counterparts and determine the $A$ and $B$  parameters by  a $\chi^2$ minimization procedure.
The scale parameter $S$ is determined iteratively by shifting the distributions so that the positions of the invariant mass peaks of the MC and data distributions agree. 

We determine the resolution parameters for three different types of muon tracks:
\label{track_smear_type}(i) tracks with SMT hits and in the full CFT acceptance ($|\etacft|<1.6$),
(ii) tracks with SMT hits and located outside the full CFT acceptance ($|\etacft|>1.6$),
and (iii) tracks without SMT hits.
\begin{figure}[!h]
\begin{center}
\includegraphics[width=0.45\textwidth]{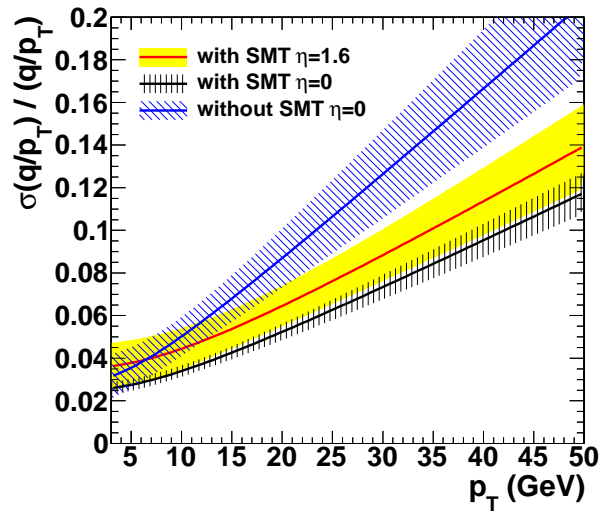}
\end{center}
\caption{ [color online] Relative muon momentum resolution 
for different types of tracks. The bands correspond to the $\pm 1$ standard deviation uncertainties on the measurements.}
\label{fig:muon_resolution}
\end{figure}

The $S$ parameter is typically around 0.5\%. 
The typical resolution parameters are given in Table~\ref{tab:resolution}~\cite{tm2540}.
\begin{table}[!htb]\center
\begin{tabular}{l c c }
\hline\hline
Track type &   $A\times 10^{3}$ (GeV$^{-1}$) & $B \times 10^{2}$ \\
\hline
SMT hits&
   $2.3 \pm 0.2 $  & $2.5 \pm 0.3 $ \\
SMT hits, $|\etacft|>1.6$&
 $2.7 \pm 0.4$& $2.2 \pm 0.7$\\
Without SMT hits&
 $4.1 \pm 0.7$&  $2.9 \pm 1.1$\\
\hline\hline
\end{tabular}
\caption{\label{tab:resolution}
Typical resolution parameters for the three types of track considered. See text for details.}
\end{table}
In these figures, the  uncertainties are assessed by increasing the muon \pt\ requirement in the $Z$ or $J/\psi$ 
selections, varying the muon and track qualities, and  varying the mass range used to compute the $\chi^2$. The statistical
uncertainties are obtained from pseudoexperiments in which the number of events in each bin of the invariant mass distributions
is varied according to its expected statistical uncertainty.
The relative momentum resolutions as a function of $\pt$ for the different types of tracks are shown in Fig.~\ref{fig:muon_resolution}. The typical resolution is 10\%--16\% for tracks of $\pt=40$~\gev.

\section{Muon backgrounds}
\label{sec:bkg}
\makeatletter{}

We identify three sources of background for physics with muons:
(i)~muons from cosmic rays,
(ii)~in-flight decays of pions or kaons into muons, and
(iii)~hadrons passing through the calorimeter (punch-through).
For high-\pt\ physics,
(iv)~real muons from the semi-leptonic decays of heavy-flavor hadrons constitute an additional source of background.
For high-\pt\ analyses, types (ii--iv) are merged under the generic name of multijet
muon background, as they all occur in events where hadrons or jets yield a reconstructed muon.
This multijet background is completely dominated by the heavy flavor component.

The contamination due to each  source of background depends on
the particular selected final state topology and on 
the identification criteria.
The numbers and examples given below are therefore only indicative.

\subsection{Cosmic rays}
The muon identification criteria described in Secs.~\ref{ssec:muid} and~\ref{ssec:track}
are optimized to reject the background composed of cosmic muons.
In particular the timing veto and the cuts on \dca\ result in a negligible contamination for most data analyses.
An example of the determination of the cosmic-ray contamination is described in Ref.~\cite{tuchming:HDR2010}
where template histograms of the \dca\ distributions are employed 
to fit to the data the contributions of cosmic rays and $p\bar p \to W\to\mu\nu$  events.
In this analysis, the template histogram for the
$ W\to\mu\nu$ signal is obtained from a selection of $Z\to\mu\mu$ events,
while the cosmic muon template is obtained from a sample with
two back-to-back reconstructed muons.
The contamination is  estimated to be below the 0.1\% level.

\subsection{Multijet background}
The multijet background is mainly reduced by the isolation criteria described in Sec.~\ref{ssec:iso}.
The multijet background can be assessed in two different ways.
Either by determining (i) the rate at which a jet yields a reconstructed isolated muon,
or  (ii) the rate at which a multijet event with an identified muon yields an isolated muon.
Once this rate is known, it can be applied in a multijet control sample to determine the contamination in the signal sample.

For case (i), the method consists of defining an unbiased background-enriched sample of multijet events and measuring the rate in that sample.
This method is employed in Ref.~\cite{Abazov:2011td} where it is found in dijet events that the probability for a jet of  $\pt>15$~GeV (100~GeV) to produce
an isolated muon of $\pt > 15$~GeV is approximately $4\times 10^{-4}$ $(2\times 10^{-3})$.

\begin{figure}[!h]
\begin{center}
\includegraphics[width=0.20\textwidth]{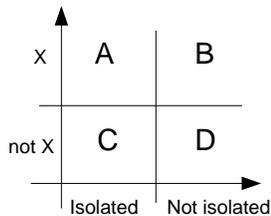}
\end{center}
\caption{ Illustration of the  ``matrix/$A B C D$'' method, where the criterion $X$, for example a requirement of high $\etmiss$,
and the muon isolation define four independent samples. See text for details}
\label{fig:abcd}
\end{figure}

For case (ii), the method defines
four different disjoint samples of events
as illustrated by Fig.~\ref{fig:abcd}.
In the first step, one selection criterion $X$ of the signal sample is reversed
to produce a multijet-enhanced sample (region $C+D$ in Fig.~\ref{fig:abcd}).
As an example, $X$ can be a high missing  transverse energy ($\etmiss$) requirement
that selects $W\to\mu\nu$ decays,
where the $\etmiss$ is obtained
from the vector sum of the transverse components
of the reconstructed muon momenta and
energy deposits in the calorimeter,
corrected for the differences in detector response of the reconstructed electrons and jets.
The rate $f$ at which a muon is isolated in that sample is then measured, $f=C/(C+D)$.
The rate $f/(1-f)$ can then be applied to a multijet control sample similar to the signal sample
but in which the isolation criterion is reversed (region $B$ in  Fig.~\ref{fig:abcd}).
Because there can be leakage of signal
into the non-isolated sample,
it may be necessary to solve a set of linear equations to
determine the composition of the different samples,
so that the method is  sometimes referred to as the ``matrix method.''
Once $f$ is properly measured, the equations read:
\begin{eqnarray}
A &=& S + EW+ MJ \\
B &=& \frac{1-\epsilon}\epsilon \times ( S + EW ) + \frac {1-f}{f} \times MJ,
\end{eqnarray}
where $S$, $EW$, and $MJ$ are the signal, electroweak background,  and  multijet background components in sample $A$, respectively, and $\epsilon$ is the isolation
efficiency for both the signal and the electroweak background.
Because the method involves defining  four independent samples of events
according to ($X$, not $X$)$\times$(isolated, not isolated) it is also known under the name of ``$A B C D$ method''.

This method is applied for example in Ref.~\cite{Abazov:2011mi},
where the selection of the $t\bar t \to \mu\nu$+jets signal events demands
$\etmiss>25$~\gev.
The isolation rate for muons in multijet events is determined with the same
selection but demanding small missing transverse energy, $\etmiss<10$~\gev;
it is found to be  17\%--22\%.

\subsection{In-flight decays and punch-throughs}

Analyses based on final states with muons from heavy flavor decays within jets may require
an accurate estimate of the in-flight decay and calorimeter punch-through contamination.
To measure the rate for kaons or pions to be reconstructed and identified as muons, a common method
consists of identifying known resonances using the invariant mass of the decay products, and  determining the rate at which
one of the decay products is  reconstructed as a muon.
For example Ref.~\cite{PhysRevD.87.072006}
uses the  $D^0 \to K \pi$  resonance and determines the rate for the kaon
to be reconstructed as a muon.
For a track requirement of two SMT hits and two CFT hits, the fraction of tracks of $\pt>5$~\gev\ originating from kaons that are reconstructed and identified as
\mediumnseg3\ muons with a converged local fit in the muon system is found to be  $(1.9\pm 0.5 ) \times 10^{-3}$.

\section{Correction to full detector simulation}
\label{sec:data_MC}
\makeatletter{}\label{sec:MC}
Physics analyses at \dzero\  widely employ the simulation of the full detector response for
standard model and  beyond the standard model processes.
Most of these processes are simulated with 
the \textsc{pythia} or \textsc{alpgen} generators, with \textsc{pythia}
providing showering and hadronization in the latter case,
followed by
a detailed {\sc geant3}-based~\cite{bib:geant} simulation of the \dzero\
detector. To model the effects of multiple $p\bar{p}$
interactions, the MC samples are overlaid with events from random
$p\bar{p}$ collisions with the same luminosity distribution as data. These events are then
reconstructed with the same software as used for the data.

Small differences are found between the data and the MC, both for the identification efficiency of muons and for their
momentum resolution. Corrections to the full \dzero\ simulation need to be applied to bring the  
simulation of the detector into better agreement with the data. These corrections are discussed below. More details
can be found in Refs.~\cite{tm2540,{tm2541}}.

\subsection{Efficiency correction}
\label{sec:efficiency_correction_data_MC}

\begin{figure*}[!]
\begin{center}
\subfigure[]
{\includegraphics[width=0.32\textwidth]{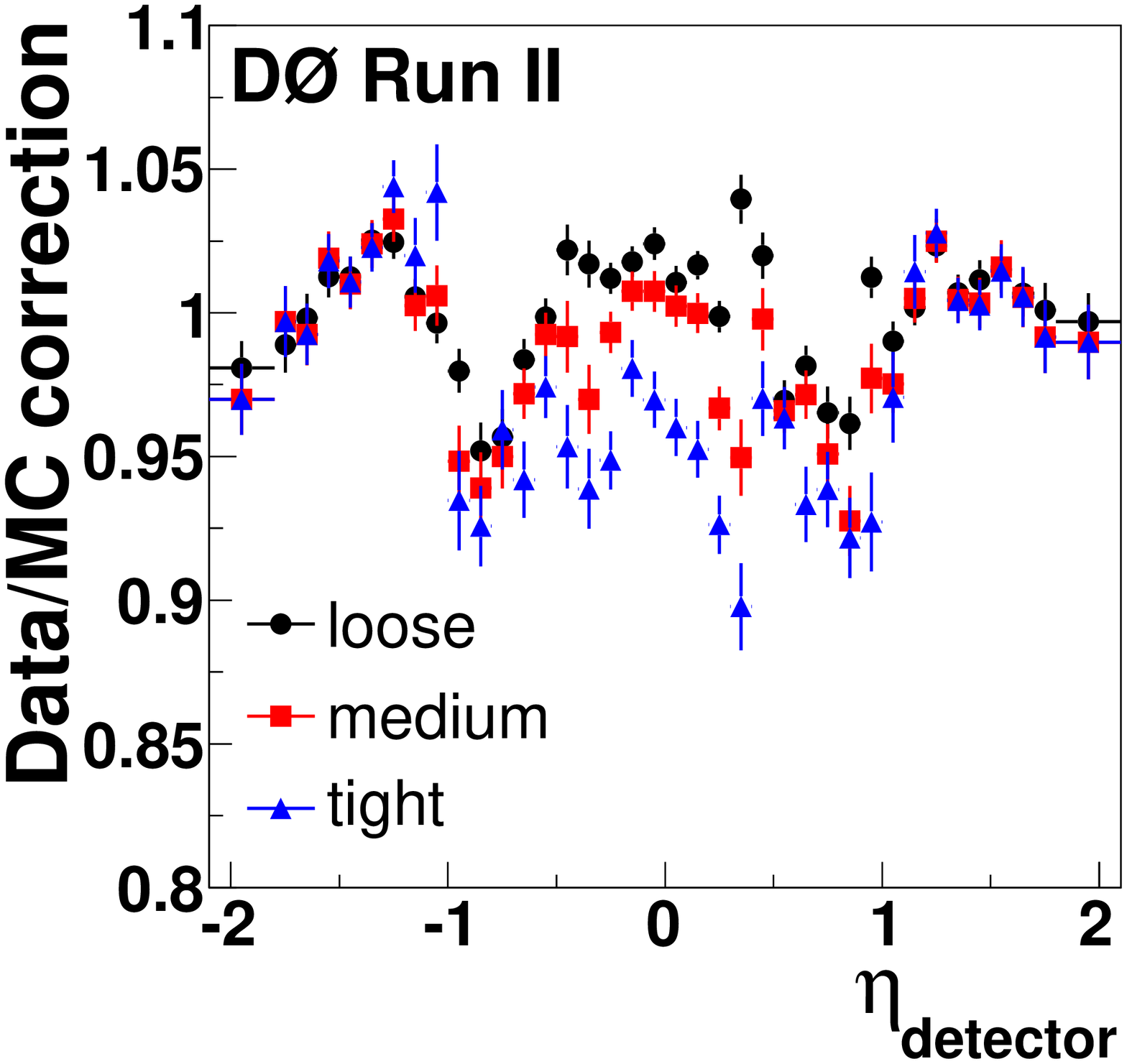}}
\subfigure[]
{\includegraphics[width=0.32\textwidth]{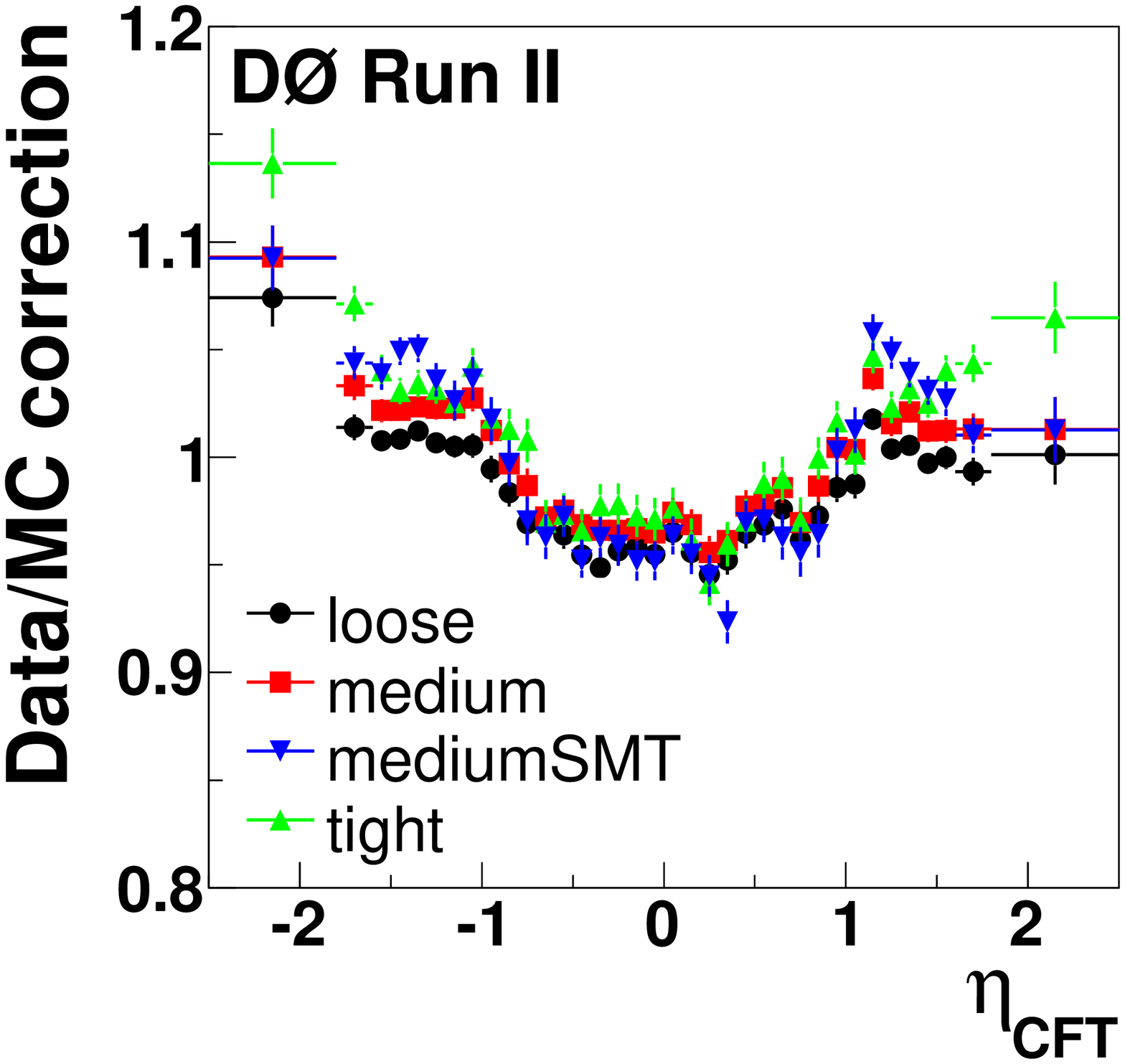}}
\subfigure[]
{\includegraphics[width=0.32\textwidth]{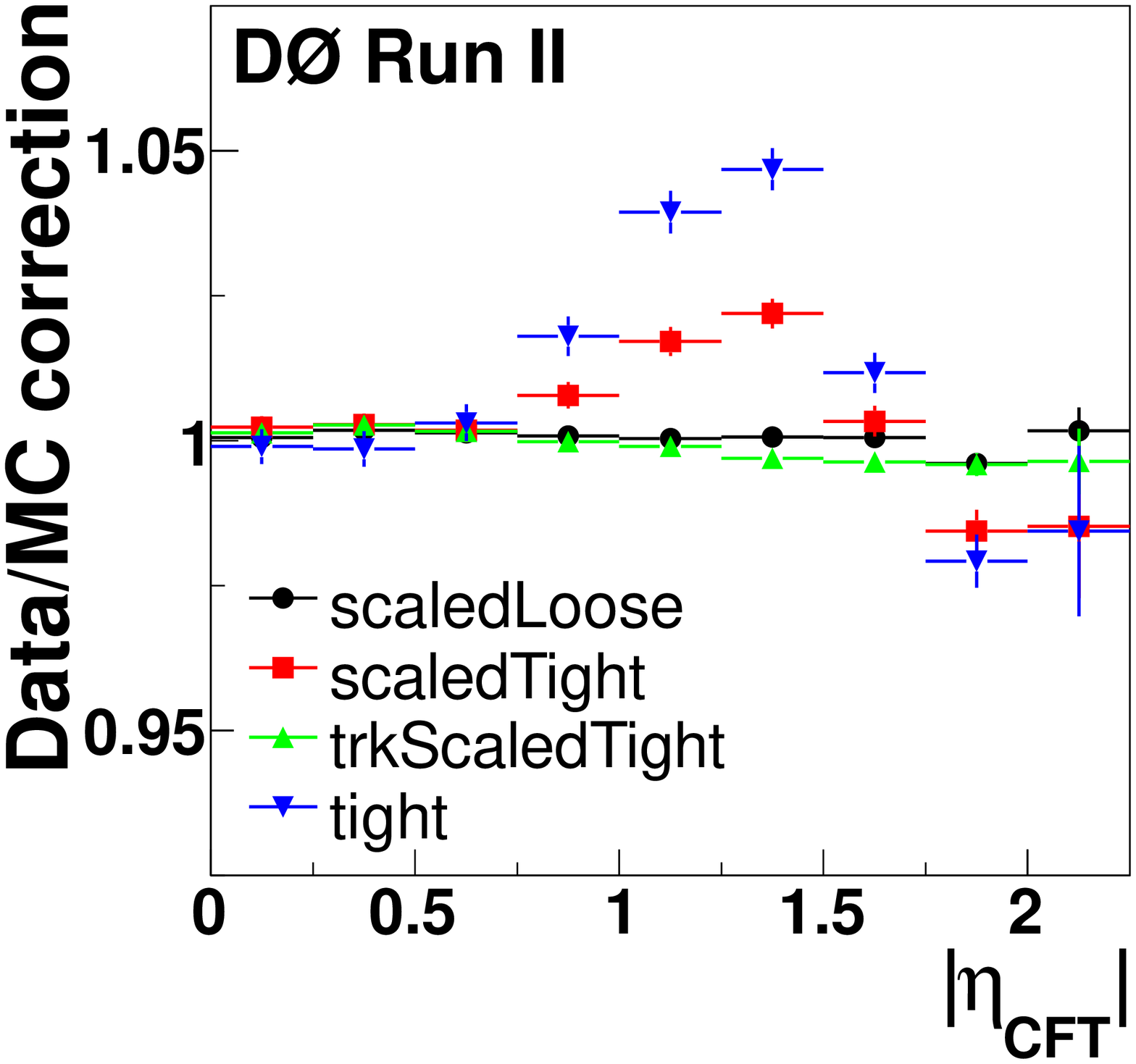}}
\end{center}
\caption{[color online] Examples of efficiency correction factors as functions of pseudorapidity for the various (a) muon-identification, (b) tracking, and (c) isolation criteria defined in Sec.~\ref{sec:identification}.}
\label{fig:efficiency_correction}
\end{figure*}

The tag-and-probe method described in Sec.~\ref{sec:performance} can be used to measure muon-related efficiencies of the full detector simulation using a \textsc{pythia} $\zmumu$ MC sample.
The same selection criteria are demanded as for the $\zmumu$  data sample, except the trigger requirement. The  luminosity spectra of the MC samples are reweighted to the same spectrum as for the measurements of efficiencies in data.

Differences are found in the measured $\zmumu$  MC  efficiencies  compared to the measured data efficiencies as can be seen in Fig.~\ref{fig:efficiency_correction}.
The relative differences are smaller than $10\%$,  and for most of the phase space they are below the 5\% level.
However, for an optimal agreement between data and simulated events
the analysis chain used by most  \dzero\ physics analyses includes weight factors to correct for these differences.
The weight factors  are obtained as the ratio of efficiencies measured in data to those measured in MC.

For the muon identification, the corrections are computed for each of the six operating points defined in Sec.~\ref{ssec:muid}.
The corrections are parametrized in the $(\etadet,\phi)$  plane
to reflect the geometry of the detector.
The dependence of the correction on luminosity is found to be negligible. On average the correction factors are 1.004, 0.988, and 0.970
for the {\em \loose},  {\em \medium}, and {\em \mediumnseg3} operating points, respectively.
The relative systematic uncertainties on these numbers are the same as the relative uncertainties for the data efficiencies given
in Sec.~\ref{sec:muon_efficiency}. They amount to 0.9\%--1.2\%.

For each of the four track reconstruction categories defined in Sec.~\ref{ssec:track}, two efficiency corrections are derived to bring MC into 
optimal agreement with data. 
The first correction accounts for the geometry of the tracking system
and is parametrized 
as a function of ($\etacft, z_0$).
The second correction accounts for the dependence of the track reconstruction efficiency on instantaneous luminosity (\lumi) for different regions in $\etacft$, and is parametrized in  ($\lumi,|\etacft|$).
The  average overall correction factors are
0.988, 1.020, 1.002, and 1.005
for the {\em \trackloose},
{\em \tracknewmedium}, {\em \tracktight}, and {\em \trackmedium}  operating points, respectively.
The systematic uncertainties on these numbers are given by the uncertainties for the measurements reported
in Sec.~\ref{sec:track_efficiency}. They amount to 1.1\%--1.6\%.

For each of the isolation operating points defined in  Table~\ref{tab:isoworkingpoints}, two efficiency corrections are applied.
All operating points demand $\dr({\rm \mu,jet})>0.5$, and 
the first correction is computed as
a function  of ($\lumi,|\etacft|$) for this requirement only. The second correction is computed
relative to the $\dr({\rm \mu,jet})>0.5$ requirement, as 
a function of $(\lumi,|\etacft|,\dr({\rm \mu,jet}))$.
This second correction is not applied for the  {\em \deltaR} operating point,
which is based solely on the  $\dr({\rm \mu,jet})>0.5$ requirement.
The  average correction for the $\dr({\rm \mu,jet})>0.5$ is 1.011.
The overall average correction factors are
1.000--1.021
for the other operating points
relative to the  $\dr({\rm \mu,jet})>0.5$ requirement,
with a relative systematic uncertainty of 0.4\%--0.8\%, dominated
by the systematic uncertainties of the measurements in data as discussed
in Sec.~\ref{sec:iso_efficiency}.
Once combined, the relative systematic uncertainty of both corrections is 0.8\%--0.9\%.

\subsection{Momentum oversmearing}
\label{sec:oversmearing}

The momentum resolution in data is typically worse by about 30\%
compared to the simulation for a muon with a transverse momentum of 40~GeV. This discrepancy reveals some mismodeling  arising from the simulation 
of hit efficiencies, the simulation of hit resolution, the magnetic field mapping,
and the internal alignment of the central tracker.

Although the simulation was constantly improved during Run~II, an ad-hoc 
smearing of the muon curvature, called oversmearing,  was developed in order to make the resolution in MC match that in data.

The oversmearing follows  Eq.~(\ref{eq:resolution_smearing}) in Sec.~\ref{sec:momentum_resolution} but instead of being applied to the true generated momentum, it is applied on
the reconstructed momentum, \ie, after the full detector simulation and reconstruction:
\begin{equation}
\frac{q}{p_T} \rightarrow (1 + S) \frac{q}{\pt} +  G   \frac{R_{\CFT}^2} { \LARM^2}  \left(A \oplus \frac{B \sqrt{\cosh \eta}}{\pt}\right).
\label{eq:smear}
\end{equation}
Here $A$, $B$, and $S$ are the oversmearing  parameters to be determined, and 
$G$ is a random number that follows a Gaussian distribution of mean zero and width one.
As in Eq.~(\ref{eq:resol2}), $\LARM$ is the radius corresponding to the outermost CFT hit of the track and  $R_{\CFT}$ is the CFT radius.
To determine $A$, $B$, and $S$, we use the same methods, the same samples  of  $\jpsimumu$ and  $\zgmumu$ events, and the same track types as in Sec.~\ref{sec:momentum_resolution}.
Typically, the $A$ parameter is determined to be around $1.7$--$3 \times 10^{-3}$~GeV$^{-1}$, the $B$ parameter around $1.4 \times 10^{-2}$, and 
the $S$ parameter around $0.3 \times 10^{-2}$.
The total uncertainty on the oversmearing parameters range from 5\% to 25\% for $A$ and $B$, and 50\% for $S$ for 
tracks of type (i) and (ii). Due to the limited statistics available to determine the parameters for type (iii), the uncertainty on the oversmearing 
reaches 100\% for this category.

Analyses requiring a good modeling of the high-\pt\ muon  momentum resolution tail use  an alternate oversmearing method.
It is similar to that described in Eq.~(\ref{eq:smear}), but we introduce a parameter $A'$ representing a resolution
term for the tail of the distribution and  a parameter $C$ representing the fraction of tracks belonging to that tail.
For a  fraction $1-C$ of randomly chosen tracks, we use the  same smearing formula as in Eq.~(\ref{eq:smear}) while  for the rest
we use the same relation as  (\ref{eq:smear}) but we replace $A$ by $A'$. This method allows 
reproduction of a double Gaussian structure in the momentum resolution function.
As for the single Gaussian case, the parameters are determined by  a $\chi^2$ minimization procedure.
The parameters $A$ and $B$ are found to be quite similar to the single Gaussian case. The parameter $C$ is determined to be around 
 3\%--8\% depending on the track type. The parameter $A'$ is determined to be around  $5$--$10 \times 10^{-3}$~GeV$^{-1}$ for types (i) and (ii),
and around  $20$--$30 \times 10^{-3}$~GeV$^{-1}$  for type (iii).
The determination of the double Gaussian parameters
suffers from higher systematic and statistical uncertainties than for the single Gaussian case. The uncertainties on $C$ and $A'$ are
at the level of 30\%--50\% for type (i) while they are  up to  $200\%$ for both types (ii) and (iii).

\begin{figure}[!ht]
\begin{center}
{\includegraphics[width=0.49\textwidth]{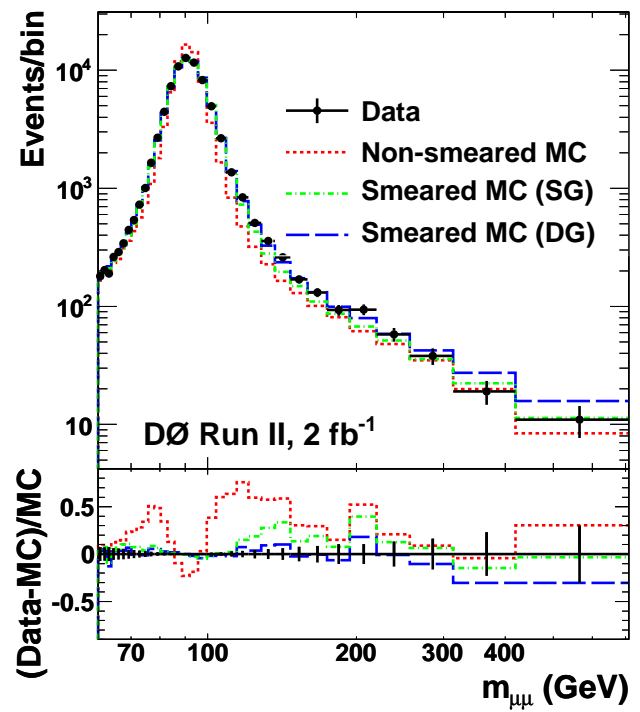}}
\end{center}
\caption{[color online] Invariant mass spectrum for high-\pt\ dimuon events, selected in the 2009--2010 data, compared to the \zgmumu\ MC before oversmearing, after oversmearing using a single gaussian function (SG), and after oversmearing using a double gaussian function (DG). See text for details.}
\label{fig:oversmearing}
\end{figure}
The necessity of the oversmearing and its effects  are illustrated in
Fig~\ref{fig:oversmearing}, where the invariant mass spectrum of  \zgmumu\ dimuon events  is compared to the MC before and after applying the oversmearing procedures.

\section{Conclusion}
\label{sec:conclusion}
\makeatletter{}

We have described the techniques and algorithms employed by the D0 collaboration during Run~II of the Fermilab Tevatron collider to
reconstruct muons from the muon system hits, and to match these muon system objects to tracks from the D0 central tracker.
We have presented the muon identification criteria employed at D0, the reconstruction and identification performances, and the experimental techniques used to measure these performances.
In the  angular region $|\eta|<2$, the muon system is able to identify
high-\pt\ muons with efficiencies ranging from 72\%  to 90\%, depending on the quality requirements.
Central tracks matched to these muons are reconstructed with efficiencies ranging from 85\% to 92\%,
depending on the quality requirements,
and a relative momentum resolution of typically 10\% for  $\pt=40~\gev$.
Isolation criteria reject multijet background for high-\pt\ physics with  efficiencies of 87\% to 99\% depending on the criteria.
Combined together these criteria have typical efficiencies of  50\% to 80\%.

The main backgrounds to the muon reconstruction have been briefly discussed.
We find that the background from cosmic rays is negligible,
while jets from multijet events yield reconstructed isolated muons with a typical probability of $4\times 10^{-4}$.

We have presented the method employed to 
optimize agreement between simulated MC and data events.
An oversmearing method corrects for the approximately 30\% difference in momentum resolution observed between the data and the default simulation.
Efficiency correction factors ranging from 0.93 to 1.02 are needed to properly simulate the efficiencies related to muon identification.

Thanks to the performance of the detector, the analysis chain, and the methods used to obtain a proper simulation of the muon reconstruction, 
the \dzero\ experiment was able to fully exploit the Tevatron Run~II data and  obtain a large number of physics results relying on muon signatures.

\section*{Acknowledgments}
\makeatletter{}We thank the staffs at Fermilab and collaborating institutions,
and acknowledge support from the
DOE and NSF (USA);
CEA and CNRS/IN2P3 (France);
MON, NRC KI and RFBR (Russia);
CNPq, FAPERJ, FAPESP and FUNDUNESP (Brazil);
DAE and DST (India);
Colciencias (Colombia);
CONACyT (Mexico);
NRF (Korea);
FOM (The Netherlands);
STFC and the Royal Society (United Kingdom);
MSMT and GACR (Czech Republic);
BMBF and DFG (Germany);
SFI (Ireland);
The Swedish Research Council (Sweden);
and
CAS and CNSF (China).

\bibliographystyle{h-physrev3.bst}
\bibliography{muonid_nim}

\end{document}